\documentclass[pre,a4paper,twocolumn,superscriptaddress,floatfix]{revtex4-1}
\usepackage[american]{babel}
\usepackage{amsmath}
\usepackage{amssymb}
\usepackage{graphicx}
\usepackage{dcolumn,bm,hyperref,color,soul,mathbbol}
\definecolor{amber}{rgb}{1.0,0.75,0.0}
\newcommand{\ud}{\mathrm{d}}
\newcommand{\ic}{\mathrm{i}}

\newcommand{\zZ}{\mathbb{Z}}

\newcommand{\old}[1]{}
\newcommand{\new}[1]{{\textcolor{black}{#1}}}

\begin{document}
\title{Ergodic and non-ergodic many-body dynamics in strongly nonlinear lattices}

\author{Dominik~Hahn}
\affiliation{Institut f\"ur Theoretische Physik, Universit\"at Regensburg, 93040 Regensburg, Germany}
\affiliation{Max Planck Institute for the Physics of Complex Systems, Noethnitzer Str. 38, 01187 Dresden, Germany} 
\author{Juan-Diego~Urbina}
\affiliation{Institut f\"ur Theoretische Physik, Universit\"at Regensburg, 93040 Regensburg, Germany}
\author{Klaus~Richter}
\affiliation{Institut f\"ur Theoretische Physik, Universit\"at Regensburg, 93040 Regensburg, Germany}
\author{R\'emy~Dubertrand}
\affiliation{Institut f\"ur Theoretische Physik, Universit\"at Regensburg, 93040 Regensburg, Germany}
\affiliation{Department of Mathematics, Physics and Electrical Engineering, Northumbria University, NE1 8ST Newcastle upon Tyne, United Kingdom}
\author{S.~L.~Sondhi}
\affiliation{Department of Physics, Princeton University, Princeton, New Jersey 08544, USA}

\date{\today}%
\begin{abstract}
\noindent
The study of nonlinear oscillator chains in classical many-body dynamics has a storied history going back to the seminal work of Fermi, Pasta, Ulam and Tsingou (FPUT). We introduce a new family of such systems which consist of chains of $N$ harmonically coupled particles with the non-linearity introduced by confining the motion of each individual particle to a box/stadium with hard walls. The stadia are arranged on a one dimensional lattice but they individually do not have to be one dimensional thus permitting the introduction of chaos already at the lattice scale. For the most part we study the case where the motion is entirely one dimensional. We find that the system exhibits a mixed phase space for any finite value of $N$. Computations of Lyapunov spectra at randomly picked phase space locations and a direct comparison between Hamiltonian evolution and phase space averages indicate that the regular regions of phase space are not significant at large system sizes. While the continuum limit of our model is itself a singular limit of the integrable sinh-Gordon theory, we do not see any evidence for the kind of non-ergodicity famously seen in the FPUT work. Finally, we examine the chain with particles confined to two dimensional stadia where the individual stadium is already chaotic, and find a much more chaotic phase space at small system sizes.
\end{abstract}
\maketitle

\section{Introduction}
The connection between Hamiltonian many-body chaos and the foundations of statistical mechanics has been an intensive 
research field for more than sixty years. Most recently, the focus has centered on the quantum setting and the highlights of this line of work include the Eigenstate Thermalization Hypothesis 
(ETH) \cite{PhysRevA.43.2046,PhysRevE.50.888,Deutsch_2018} and the complementary discovery of absence of thermalization in many-body 
localized systems \cite{PhysRevB.76.052203,PhysRevLett.95.206603,PhysRevB.75.155111, RevModPhys.91.021001}.  

An important role on the classical side has been played by studies of one dimensional mass-spring systems or oscillator chains with anharmonicities. The purely harmonic chains are, of course, integrable, and their normal modes give rise to an extensive set of conserved quantities. The challenge has been to add anharmonicities or non-linearities and to see ergodic behavior emerge. Indeed, one of the most celebrated parts of this body of work is the Fermi-Pasta-Ulam-Tsingsou (FPUT) problem
 \cite{osti_4376203,FORD1992271,doi:10.1063/1.1855036} whose identification is really what started off the field in the first case. As is well known, eponymous authors intended to analyze the energy sharing among the normal modes in a perturbed linear chain with weak cubic or quartic anharmonicities taking advantage of the newly developed computers. To their surprise, instead of equipartition the system showed signatures of recurrences even 
after long times. 

The resulting investigations led to both an understanding of this phenomenon and of its limitations---during a period of explosive growth in our understanding of nonlinear dynamics in classical systems and the phenomenon of chaos. Here we should flag the work of Chirikov and Izrailev who identified an energy separating the non-ergodic behavior found by FPUT from ergodic motion, based on the resonance-overlap criterion \cite{1966SPhD...11...30I}. 
For sufficiently small nonlinear interactions the resonances of the associated perturbations do not overlap 
so that chaotic layers stay constrained to small phase space regions. 
When neighboring resonances overlap the chaotic layers can spread over the entire phase space leading to a enhanced
energy sharing among different normal modes. Indeed, 
FPUT had suggested, in more modern language, that the critical energy density required for resonance overlap vanishes in the limit of large particle numbers, such that equipartition is obtained in the thermodynamic limit \cite{doi:10.1063/1.1855036}.

The understanding of the low energy regime came from looking at the continuum limit of FPUT \new{for specific initial conditions} and finding that it is the integrable \old{modified}\new{(normal or modified)} Korteweg de Vries equation that, {\em e.g.}, gives rise to the formation of solitons \cite{PhysRevLett.15.240}, see also \cite{benettin2013fermi} for a recent \old{review of the connection with integrable PDEs}\new{connection for more generic initial conditions with integrable systems}. 
The analysis of the latter and related models has also given deeper insight into the role of stable phase space islands 
for the global nonlinear dynamics, see {\em e.g.} about discrete breathers in \cite{mackay2000discrete,FLACH1998181} and references therein.\\

In this paper, we introduce a new family of nonlinear oscillator chains and initiate their study. We are motivated by two objectives. First, these models have a degree of tractability as they involve linear time evolution interrupted by instantaneous non-linearities\new{, in a similar fashion as Chirikov's standard map (aka kicked rotor) for low-dimensional chaos, see {\em e.g.} \cite{Chirikov:2008}}. Second, they provide an interesting point of departure for examining the nature of the phase space in the infinite volume limit as they permit us to introduce a high degree of chaos already at the level of a {\it single} degree of freedom. This creates the possibility that we will observe clear signatures of many-body chaos for relatively modest numbers of degrees of freedom even in the interacting system.

\begin{figure}[!ht]
	\centering
	\includegraphics[width=0.8\linewidth]{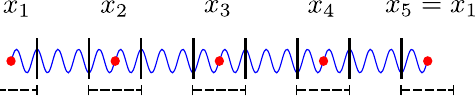}
	\caption{Sketch of our system. Each site obeys billiard constraint while it is connected via a harmonic potential to its nearest neighbor. Periodic boundary conditions are assumed at the ends of the chain. }
	\label{sketch}
\end{figure}

The models are easily described: we arrange a set of $n-$dimensional stadia/billiard tables/domains with hard walls on a $d$ dimensional regular lattice and populate each with a single particle. Then we couple the particles with harmonic springs. From the viewpoint of many-body physics, these systems fall in the class of discretized classical field theories with $n$ independent fields in $d$ space dimensions.
It may be worth mentioning that, conversely to the FPUT problem, our choice of onsite constraint breaks the translation invariance, and the total momentum is not conserved anymore. Depending on the choice of geometry for the stadia, we can build in various internal symmetry groups and we expect equilibrium computations on these models to show the same qualitative behavior as their more familiar relatives, e.g. multi-component Landau-Ginzburg-Wilson models with quartic interactions. In this paper we will study examples in $d=1$ with $n=1$ with a $\zZ_2$ symmetry (see Fig.~\ref{sketch}) and $n=2$ with $\zZ_2 \times \zZ_2$ symmetry.

From the viewpoint of single-particle chaos theory our models 
immediately connect to the study of single-particle classical and quantum chaos in hard-wall billiard 
systems~\cite{balian1970distribution,balian1971distribution,balian1972distribution,gutzwiller1991chaos,10.5555/1214825,Stoeckmann}
such as the stadium billiard~\cite{bunimovich1979} which are among the simplest systems to exhibit chaotic dynamics. 
Indeed, different shapes of the billiard \cite{Robnik_1985} or additionally applied external fields 
\cite{Wintgen-Freidrich-1989,RICHTER19961} lead to integrable, weakly or strongly chaotic classical dynamics. The most common case is that the Hamiltonian dynamics is not fully ergodic 
but characterized by a mixed phase space where locally integrable or near-integrable dynamics coexist with regions governed by unstable hyperbolic dynamics \cite{Bohigas-et-al-1993,bunimovich2001mushrooms}. Hence we see that our models allow for substantial chaos to be built in at the lattice scale as advertised above.
There could be also some adiabatic invariant slowing down the thermalization significantly as recently investigated in \cite{iubini2019dynamical}.

\noindent

In the balance of the paper we begin by more formally defining our models in Sec.~\ref{model}. Next we study the phase space for $n=1$ (scalar field at each site), and $N=2$ particles via Poincar\'e surfaces of section
 to get a sense of the dynamics. The main regular regions (stable islands for such a low dimensional case) are identified, together with the central periodic orbit. For larger values of $N$ the whole Lyapunov spectrum is first analyzed for arbitrary initial conditions.
In the thermodynamic limit the positive part of the Lyapunov spectrum converges numerically to a smooth curve 
and shows no vanishing Lyapunov exponents (up to {one} corresponding to the energy conservation)
for the choice of initial conditions in the chaotic sea, see Sec.~\ref{phasespace}.
Further we consider special initial conditions, which generalize the stable islands seen for $N=2$.
In Sec.\ref{stabilitycontinuum} we consider smooth initial configurations where the chain starts as a rigid bar. While the continuum limit is integrable, we observe energy sharing among normal modes, caused by the singular confinement potential.
Also we probe the short wavelength limit by analyzing {the excitations of a single particle.}
It turns out that the confinement potential suppresses energy sharing among different particles in this limit.
In Sec.~\ref{recovery of statistical mechanics}, we compare the results for 
two-particle correlation functions obtained by the canonical ensemble, and by molecular dynamics: we can see a very good agreement reinforcing the idea that most of the phase space is chaotic. Nevertheless
small deviations between both results imply the existence of {invariant phase space regions where the dynamics is locally integrable}. 
In Sec.\ref{stadium_results} we introduce and briefly discuss the model for $n=2$, for which there can be chaos at each site of the lattice.

\section{Models} \label{model}

\subsection{$n=1$, $\zZ_2-$symmetric chain}

We present here two possible ways to define our model: a discretized field theory with periodic boundary conditions or a closed chain.
First consider a discretized field theory on a lattice of $N$ sites with a unit mesh, $\varphi_i$ denotes the value of the field at the $i$th site.
The Lagrangian for the field is
\begin{equation}
  \mathcal{L}=\frac{1}{2} \sum_{i=1}^N  m \dot{\varphi_i}^2 - \frac{k}{2}\sum_{i=1}^N (\varphi_{i+1}-\varphi_i)^2 - V(\varphi_i)\ ,  \label{Lagrangian}
\end{equation}
where $m$ (resp. $k$) are the mass (resp. spring constant) of the field. Note that periodic boundary conditions are assumed.
The local, or on site, potential $V(\varphi)$ is taken to mimic the presence of hard walls:
\begin{align}
\begin{split}
V(\varphi) = \begin{cases}
0 & \,0<\varphi<\varphi_0 \, , \\
\infty  & \, \text{otherwise} \, ,
\end{cases}\\ 
\end{split}
\end{align}
Rescaling the field via $\tilde{\varphi_i}=\varphi_i/\varphi_0$, and the time via $\tilde{t}=\sqrt{k/m}\ t$, the Lagrangian (\ref{Lagrangian}) can be rewritten as, after dropping the tildes:
\begin{equation*}
  \mathcal{L}=\frac{k\varphi_0^2}{2}\left[\sum_{i=1}^N  \left(\frac{\ud \varphi_i}{\ud t}\right)^2 - \sum_{i=1}^N (\varphi_{i+1}-\varphi_i)^2 - V(\varphi_i)\right]
\end{equation*}
Later we shall rather use the Hamiltonian formulation, and measure the energy in units of $k\varphi_0^2/2$. Relabeling the field value as $x_i$, and the corresponding momentum as $p_i$, the final Hamiltonian is:
\begin{equation}
    H=\sum_{i=1}^{N} p_i^2+(x_{i+1}-x_i)^2+V(x_i).\label{bft_scal}
\end{equation}
with the potential
\begin{align}
\begin{split}
V(x) = \begin{cases}
0 & \,0<x<1 \, , \\
\infty  & \, \text{otherwise} \, ,
\end{cases}\\ 
\end{split}
\end{align}
{\em i.e.}, if a particle hits the wall, the sign of its incoming momentum is reversed. The Hamiltonian (\ref{bft_scal}) will be the central object in our study. We can arrive at our field potential as the limit $\lim_{p \rightarrow \infty} (x-1/2)^{2 p}$. The choice $p=2$ yields the standard interacting Klein-Gordon field in $d=1$ which has been studied at length previously \cite{aarts2000exact,boyanovsky2004approach,danieli2019dynamical,kevrekidis2019dynamical}. Our model is also similar to a model used for DNA denaturation, where the hard wall limit appears as the singular limit of the exponential damping of the interaction term \cite{peyrard1989statistical}.

The second way of introducing our model is to consider a closed {chain} of $N$ classical particles with harmonic nearest-neighbor interactions, each moving in one-dimension while sitting
in a box of length $L=Na$, where $a$ is the distance between two particles at rest. 
When one is interested in the variations around the equilibrium, it is relevant to introduce the deviation of the position of each particle from its rest position: $\varphi_i=x_i-a$
The Lagrangian for such a closed chain is
\begin{equation}
  \mathcal{L}=\frac{1}{2} \sum_{i=1}^N  m \dot{\varphi_i}^2 - \frac{k}{2}\sum_{i=1}^N (a+\varphi_{i+1}-\varphi_i)^2 - V(\varphi_i)\ ,
\end{equation}
with the same definition for $m$, and $k$ as above. The key ingredient of our model is in the choice for the local potential
\begin{align}
\begin{split}
V(\varphi) = \begin{cases}
0 & \,0<\varphi<a \, , \\
\infty  & \, \text{otherwise} \, ,
\end{cases}\\ 
\end{split}
\end{align}
Due to our choice of a closed chain, {\em i.e.} periodic boundary conditions, only the quadratic term remains in the interaction part. Performing the change of variable $\tilde{\varphi_i}=\varphi_i/a$, and $\tilde{t}=\sqrt{k/m}\ t$, leads again to the Hamiltonian (\ref{bft_scal}) through the above mentioned steps.

The Hamiltonian (\ref{bft_scal}) is invariant under the map, which reverts every position as
$$ x_i \mapsto 1-x_i \ .$$
As it is an involution it gives rises to a $\zZ_2-$symmetry.
The Hamilton equations of motion are
\begin{align}\label{EOM}
\begin{split}
&\dot{x_i} = 2p_i \\
&\dot{p_i} = 2 (x_{i+1}-2 x_i+x_{i-1}).
\end{split}
\end{align}
up to reflections at the walls.

\subsection{$n=2$, $\zZ_2\times \zZ_2-$symmetric chain}

\begin{figure}[!ht]
  \centering
  \includegraphics[width=0.8\linewidth]{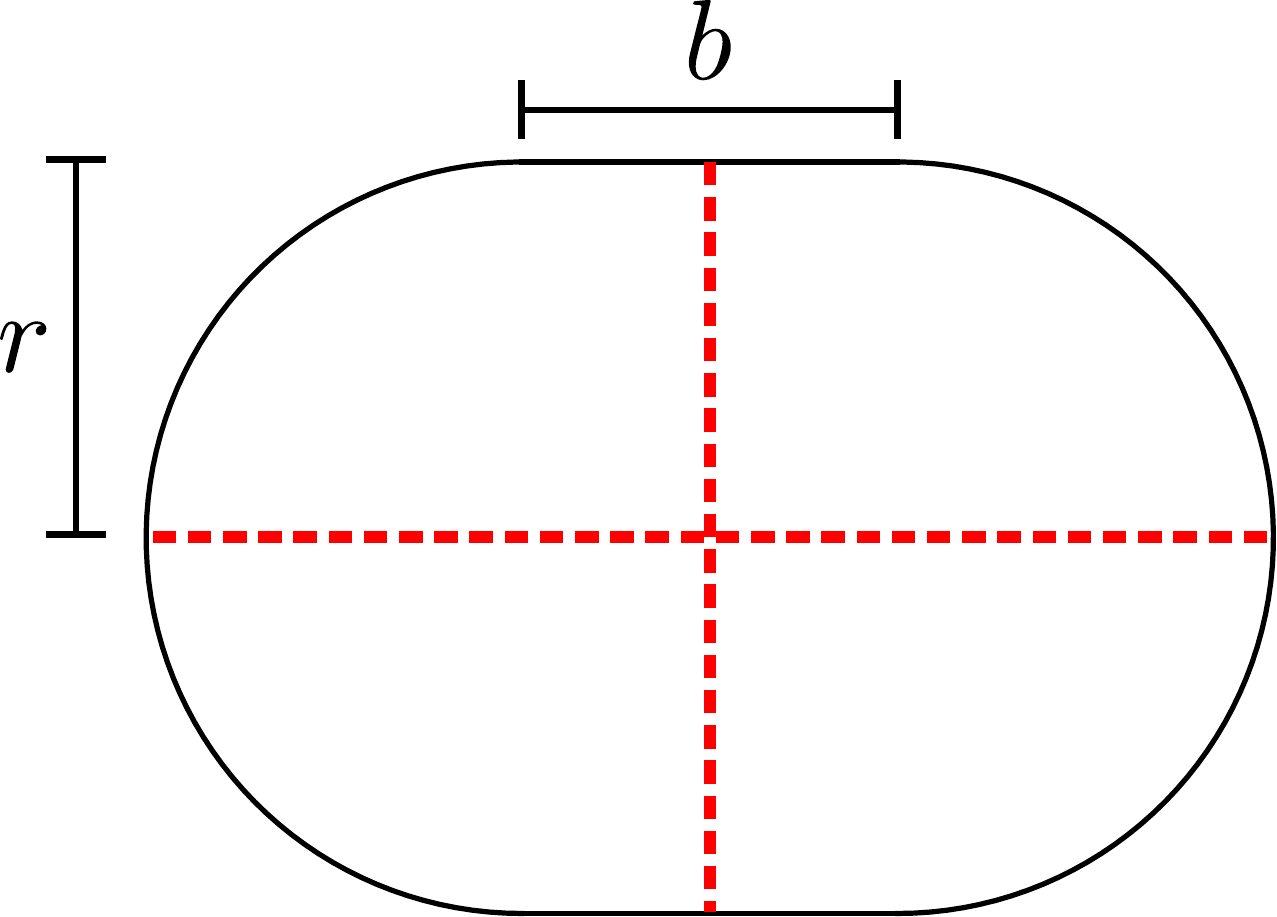}
  \caption{A sketch of the stadium billiard. For later purposes, we use $r=0.5$ and $b=0.5$. The lines stand for the symmetry axes.}
  \label{billiard}
\end{figure}

The introduced model can be easily generalized to multi-component scalar fields, which allows for arbitrary, and a larger variety of confining geometries.
For later use, we introduce a doublet of scalar fields in a stadium billiard:
\begin{equation}
    H=\sum_{i=1}^{N} {p_x}_i^2+{p_y}_i^2+(x_{i+1}-x_i)^2+(y_{i+1}-y_i)^2+V(x_i,y_i),\label{bft_vect}
\end{equation}
where $V(x,y)$ is the confinement potential for a Bunimovich stadium billiard~\cite{bunimovich1979} with $r=0.5$ and $b=0.5$, see Fig.~\ref{billiard}. In particular, when hitting the wall the linear combination of ${p_x}_i$ and ${p_y}_i$ giving the momentum in the normal direction is reversed, whereas the linear combination defining the tangential momentum is conserved. In that setting, even the single-particle motion is chaotic. We will use that particular geometry in Sec.~\ref{stadium_results}.

As the Hamiltonians (\ref{bft_scal}) and (\ref{bft_vect}), and the constraints are time-independent, the total energy $E$ is a constant of motion. Hence $E$ and $N$ remain as the only free parameters
of the problem, and the energy density $h=E/N$ is used as relevant control parameter. This choice of scaling is different from the recent study of the largest Lyapunov exponent as reported in FPUT problem \cite{mulansky2011strong}.
For later reference we further introduce the frequencies of the normal modes
\begin{align}
\omega_{i}=4\sin \left({ i \frac{\pi}{N}}\right), \quad 0\leq i\leq N-1 
\end{align}
of the free problem, {\em i.e.} $V(x_i) \!=\! 0$.

\subsection{Limiting cases}
Before we present and analyze our numerical simulations we consider two relevant limiting cases with respect to
the energy density.
Due to the presence of the walls the maximum scaled distance between any two particles is smaller than $1$,
and hence the interaction energy is bounded (by $N$). 
For energy densities $h \gg 1$, the minimum kinetic energy per particle is therefore $e_{\text{kin}}=h-1$, 
leading to a minimum momentum per particle $p_{\text{min}}=\sqrt{e_{\text{kin}}}=\sqrt{h-1}$. 
If in the regime $h \gg 1$  (or $p_{\text{min}} \gg 1$) the effect of the interaction between different 
particles is neglected, the system reduces to $N$ independent particles in a billiard. This regime seems as the most favorable to see the recently introduced glassy dynamics \cite{danieli2019dynamical}, even if we did not investigate it specifically.
On the contrary, the limit $h \ll 1$ resembles a tight, nearly free harmonic chain, where the energies of the 
individual normal modes are redistributed from time to time due to infrequent collisions with the walls. 
As shown below, in this case, the dynamics of the system is weakly chaotic still leading to an information loss 
of the initial configuration.
These two extreme regimes of $h$ are roughly separated at $h \approx 1$ that is expected to be the most chaotic regime.
We finally note that for $h>1$ the entire configuration space is accessible due to the upper bound for the interaction.

\section{Phase space analysis for the scalar ($n=1$) model }\label{phasespace}
\subsection{Two particles - Poincar\'e surface of sections}

It is instructive to start with two particles ($N=2$), each having only one degree of freedom ($n=1$), since the underlying dynamics is still easy to visualize.
The Hamiltonian reads
\begin{equation}
    H_2=p_1^2+p_2^2+(x_2-x_1)^2, \label{H2}
\end{equation}
where $0\leq x_1,{}x_2\leq1.$

\begin{figure}[ht]
	\centering
	\includegraphics[width=\linewidth]{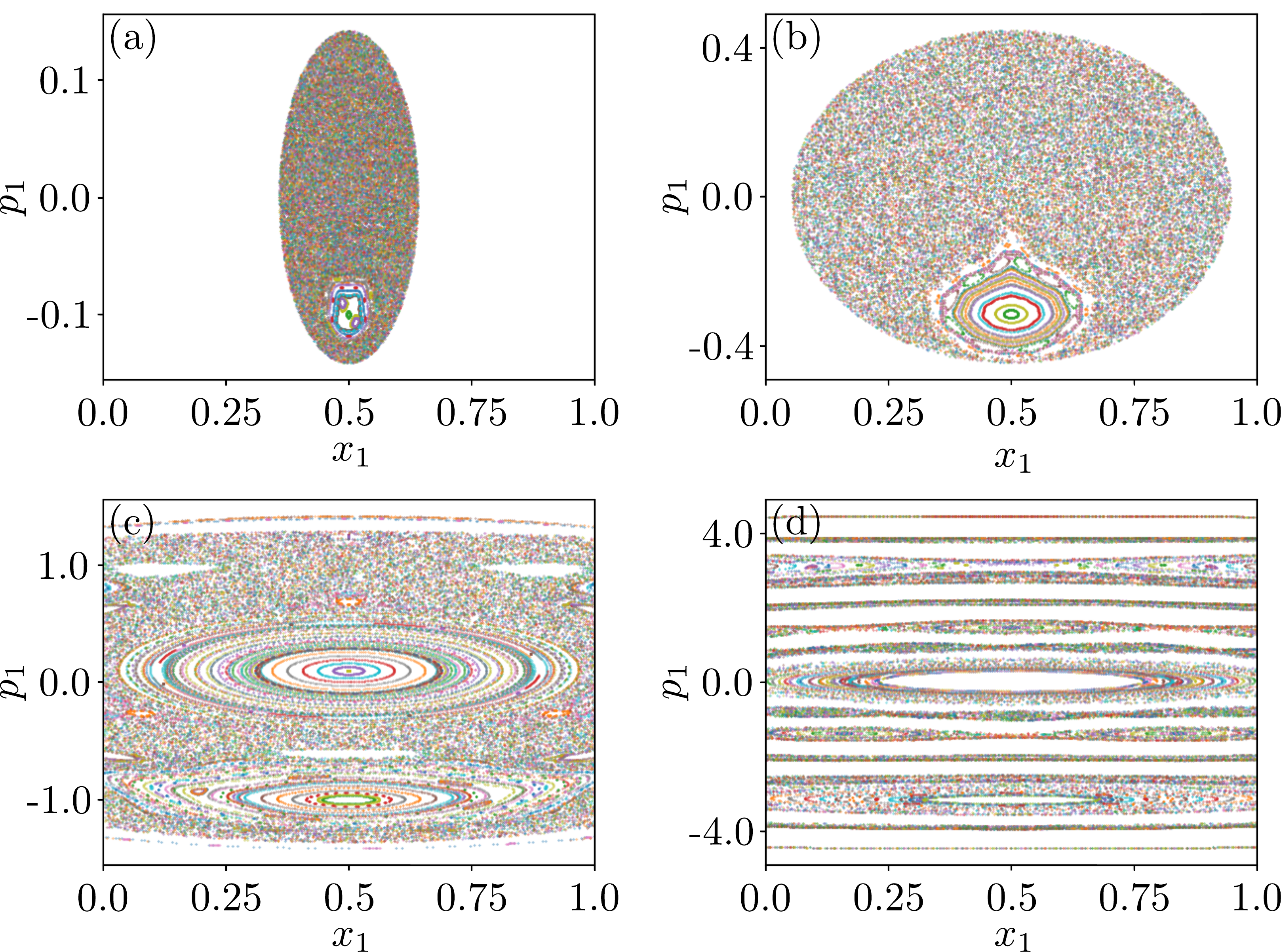}
	\caption{Poincar\'e surfaces of section visualizing the $N=2$ particle dynamics induced by the Hamiltonian (\ref{H2})
for energy density (a) $h=0.01$, (b) $h=0.1$, (c) $h=1$ and (d) $h=10$. Different colors mark different trajectories.
About $250$ initial conditions are used by choosing $x_1(0)$ and $p_1(0)$ on a regular $16 \times 16 $ grid 
on the accessible phase space with $x_2(0)=0.5$ and $p_2(0)<0$.}
	\label{fig:Poincared}
\end{figure}
\noindent
Figure \ref{fig:Poincared} shows Poincar\'e surfaces of section (PSoS) of the corresponding phase space:
the canonical coordinates $x_1(t)$ and $p_1(t)$ of the first particle are plotted at each time $t$ when the second particle
reaches the symmetry point $x_2 = 0.5$ with momentum $p_2(t)<0$.

The system exhibits clear features of integrable dynamics for large $h$, as visible in panel d):
The PSoS displays cuts through tori in phase space, each creating a single quasi-smooth curve. 
This regime of large $h$ can be seen as a perturbation of the non-interacting case where the PSoS would simply consist of horizontal lines.
Conversely, at small or moderate $h$, See panels a),b) in Fig.~\ref{fig:Poincared}, larger regions of the PSoS appear uniformly filled, 
indicating ergodic dynamics.

At intermediate energy density $h=1$, see panel c), the phase space is dominated by two large stable islands centered around
two fixed points. Note that the lower one persists even in the limit of small $h$.
Indeed, for all energy densities, there exists a stable fixed point at $x_1(0)=x_2(0)=0.5$, $p_1(0)=p_2(0)$. 
The positions of both particles coincide, which minimizes the interaction, and they move together 
as one single entity.
It is easy to determine that this corresponds indeed to $x_1(0)=x_2(0)=0.5$, and $p_1(0)=p_2(0)=\sqrt{h}$, as is illustrated also in Fig.~\ref{fig:Poincared}.
We will later consider the generalization of this fixed point for a chain of $N$ particles. The vicinity of this phase space point can be then described by a smooth continuum limit.
For energy densities $h \gtrsim 1$, the second stable island emerges at $x_1(0)=0.5$ and 
$p_1(0) \rightarrow 0$ for $h \rightarrow \infty$ \cite{report}. 
While in this case the second particle bounces rapidly off the walls, the position of the first particle
at the center of the box is only slightly disturbed. Again we will detail the generalization of such {an excitation of a single particle} or driven motion. 

\subsection{$N$ particles and general initial conditions} \label{stability}

For a larger number of particles it becomes quickly prohibitive to probe the entire phase space with a narrow grid 
of initial conditions. 
Instead, we calculated the Lyapunov spectra for $M=100$ different, randomly chosen initial conditions for various
given total energies in order to explore how ergodic the phase space dynamics is for different $h$.

\begin{figure}[ht]
  \includegraphics[width=0.65\linewidth]{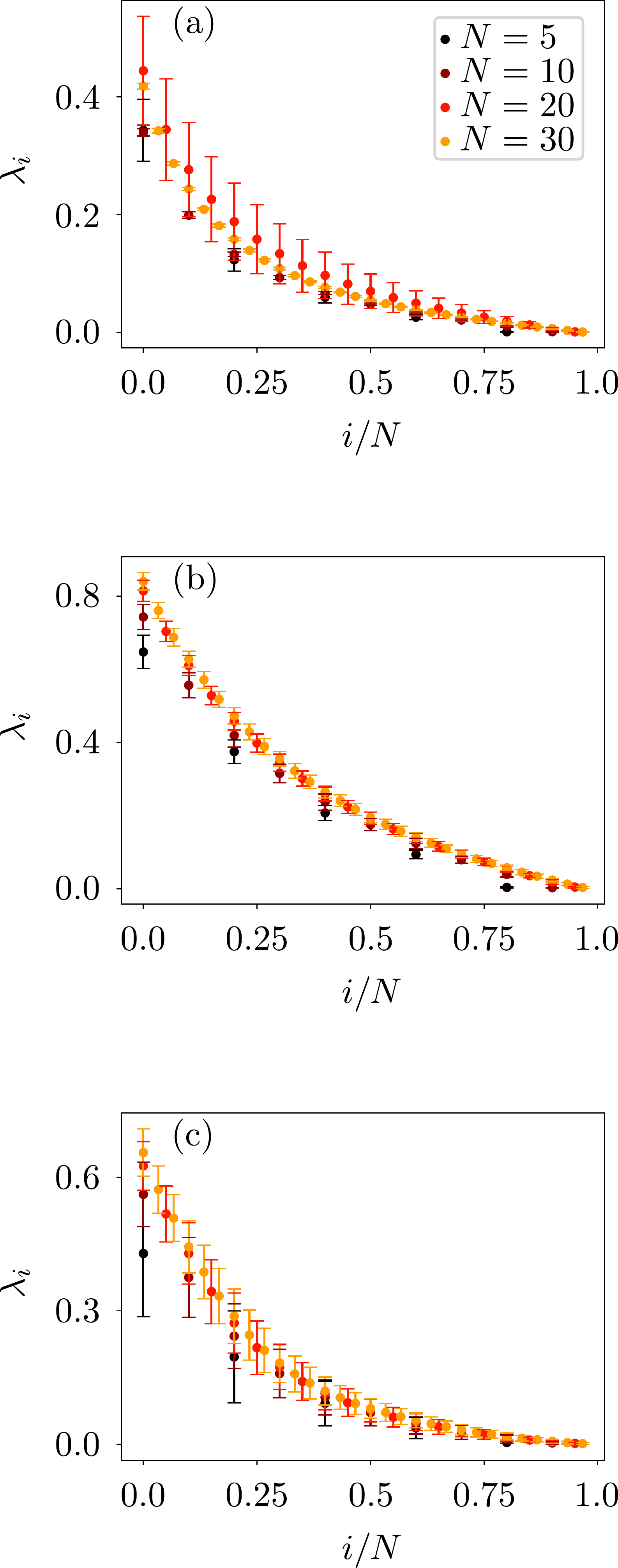}
  \caption{\new{Positive part of the Lyapunov spectrum, as a function of their (rescaled) index, for different particle numbers $N$ and different energy densities (a) $h= 0.1$, (b) $h= 1$ and (c) $h= 10$. The maximal Lyapunov exponent (for $i= 1$) reaches a maximum when h varies around $h\sim 1$.}}
  \label{fig:Lyapunovspectra}
\end{figure}

\noindent
The exponential divergence in time of two neighboring trajectories starting with a small deviation $\delta \Gamma(0)$ from
an initial point $\Gamma(0)$ in phase space is quantified through the maximal Lyapunov exponent 
that is defined as
\begin{equation}
  \lambda_{\Gamma(0),{}\delta \Gamma(0)}=
  \lim_{t\rightarrow \infty} \frac{1}{t} \ln \frac{|\delta \Gamma(t)|}{|\delta \Gamma(0)|}\, .
\end{equation} 
We numerically compute this exponential growth rate in the $2N$-dimensional phase space.
For each initial condition, there exist $2N$ different Lyapunov exponents, which we sort in decreasing order, 
$\lambda_1>\ldots \lambda_{2N}$, see e.g. \cite{PhysRevA.14.2338} for an implementation procedure.
In a closed Hamiltonian system the symplectic structure of the dynamical map implies that the phase space volume 
is conserved and the Lyapunov exponents are connected via the pairing rule  $\lambda_i=-\lambda_{2N-i}$, 
see e.g.~\cite{AbrahamMarsden}. Hence it is sufficient to consider only the first half of those exponents.
To compute the Lyapunov exponents we followed the trajectories for a time range up to $10^5$ collisions,
depending on energy density $h$ and particle number $N$, until a convergence within $3\%$ of relative accuracy was achieved for
the positive Lyapunov exponents. 

The resulting Lyapunov spectra are depicted in Fig. \ref{fig:Lyapunovspectra}. The panels (a) to (c) show,
for increasing energy density, $N-1$ positive Lyapunov exponents, so there are no global integrals of motion save
the total energy. 
The different spectra in each panel exhibit  convergence towards continuous curves with increasing $N$ 
as the number $N\!-\!1$ of points increase and the error bars shrink. 
Note that the fact that one finds $N\!-\!1$ positive Lyapunov exponents does not preclude the presence of 
stable regions in phase space. 
{The numerical convergence towards a continuous curve} was already made for the standard FPUT-chain \cite{Livi_1986}, a three-dimensional 
Lennard-Jones potential \cite{Posch,MORRISS1989307,PhysRevA.38.473} and for a hard sphere gas \cite{DELLAGO199768}. 
When considering a larger number of initial conditions ({\em i.e.} a finer grid of initial points on the constant 
energy surface), the errors bars get smaller hence a better convergence towards a smooth curve. This convergence of the Lyapunov spectra towards a continuous curve indicates a dominant phase space region of unstable motion {\em i.e.} chaotic dynamics.
While we see a clear decrease of the smallest Lyapunov exponent for increasing $N$ at any value of $h$, our numerics do not enable us to draw a definite conclusion about the limit.

\subsection{Regular regions in phase space}
\label{stabilitycontinuum}

In this Section we describe more precisely three different families of so called regular initial conditions, {\em i.e.} they may lead to a non ergodic long time behavior (hence failure of thermalization). The goal is first to emphasize the mixed character of the many-body phase space. Second we discuss the size of some families of regular initial conditions, to understand how large their contribution is when going to the infinite size limit. Two of the families generalize the two stable islands we identified for $N=2$. But the third is new to the case of large $N$.

\subsubsection{Near-uniform motion}

In order to explore whether stable regular phase space regions exist in the large-$N$ limit, we begin with the
many-particle generalizations of the two-particle fixed point {$(x_1(0),x_2(0),p_1(0),p_2(0))=(0.5,0.5,p_0,p_0)$} with $p_0<0$ 
in Fig. \ref{fig:Poincared}. 
For more than two particles, it is given by the conditions $p_i(0)=p_j(0),\,x_i(0)=x_j(0)$ 
for any $\,i,j \in \{1 \dots N\}$. It corresponds to a common motion of all particles sitting at the same position 
as a "rigid body". One should stress that the rigid motion exists at {\em any} given value of the total energy.

\begin{figure}[!ht]
  \centering
  \includegraphics[width=0.95\linewidth]{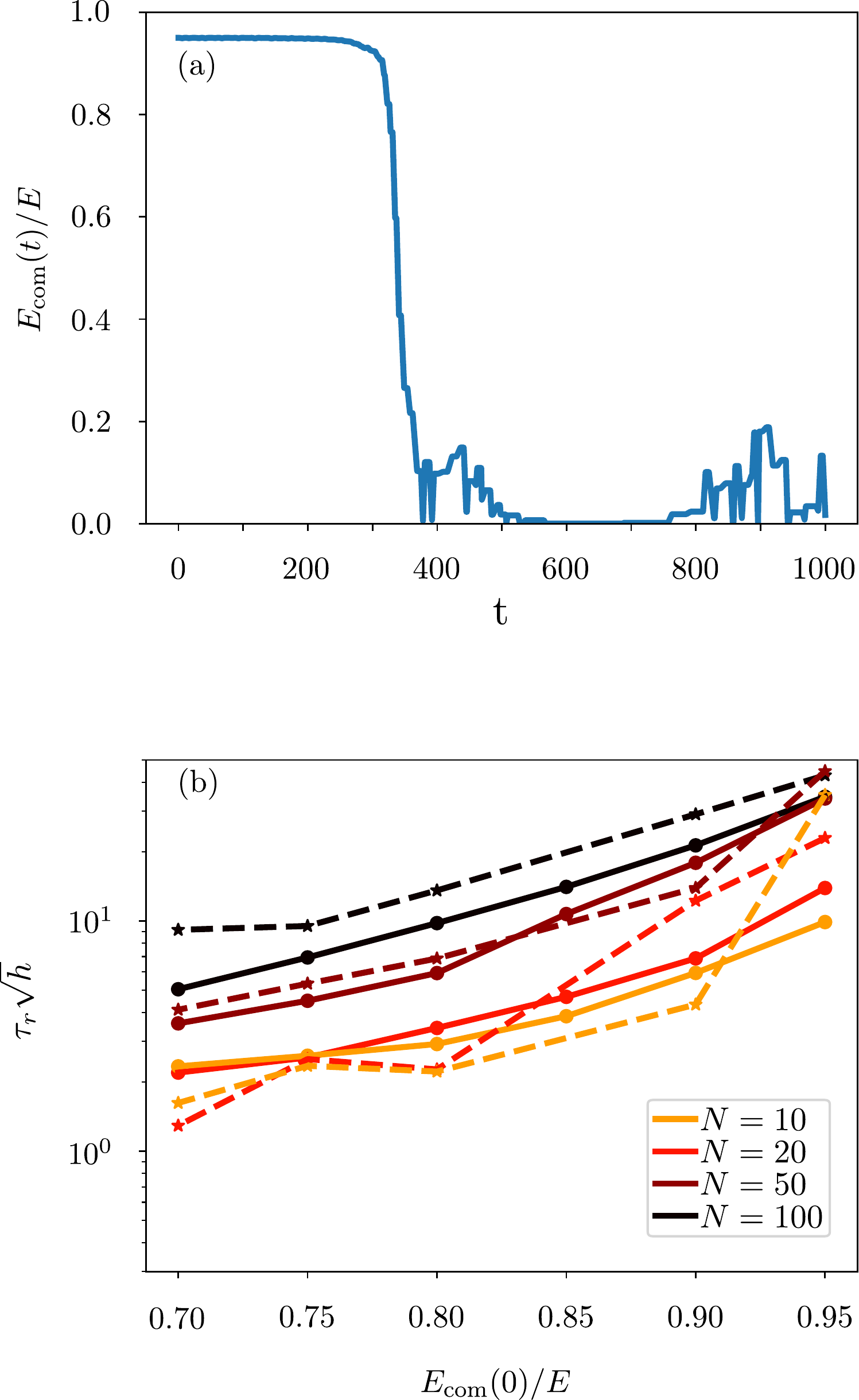}
  
  \caption{
\new{(a) Typical time evolution of $E_{\text{com}}(t)$, Eq.~(\ref{eq:ECOM}), for a trajectory near the stable $N-$particle fixed point with initial conditions $E_{com}(0)/E=0.95$ and $h=0.01$. One can detect the relaxation of the center-of-mass mode after a time $t\sim 400$, this corresponds to roughly $80$ subsequent reflections on the wall. (To get a smoother curve the selected times are those for which no particle is too close to the walls.) (b) Dependence of the relaxation time $\tau_r$ on the ratio $E_{com}(0)/E$ for different particle numbers. Dashed lines: $h= 0.1$, Full lines: $h= 0.01$. }
}
  \label{fig:fixed point}
\end{figure}

To analyze the stability behavior of this fixed point we studied the properties of trajectories in its vicinity.
To this end we considered trajectories with an excitation of the first normal mode in analogy to the original 
FPUT setting and introduce, as a convenient measure, the time-dependent center-of-mass (COM) energy 
\begin{equation}
    E_{\text{com}}(t)=\frac{1}{N} \left(\sum_{i=1} ^N p_i(t)\right)^2 \, .
\label{eq:ECOM}
\end{equation}
The "rigid-body fixed point" {is the only phase space point obeying} the condition $E_{\text{com}}(0)/E\!=\! 1$. \new{This can be proven shortly as follows. Certainly this fixed point obeys this condition. Now assume this condition is obeyed at a certain time for a given trajectory. This means in particular that all the particles sit at the same position inside their respective box (otherwise the potential energy would be nonzero). If no particle reaches the wall, then the center of mass energy is a conserved quantity (because the evolution is free) and the condition is satisfied.
If a particle reaches a wall, so do all the others because they sit at the same position inside their box. This means that all the momenta are simultaneously reversed: $p_i\mapsto \ -p_i$. The center of mass momentum has then also its sign reversed: $\sum_i p_i\mapsto \ -\sum_i p_i$. The energy of the center of mass remains unchanged, and the condition $E_{com}/E=1$ again holds after this collision.}
The time evolution of ${E_{\text{com}}(t)}$ for a typical trajectory of $N=100$ particles near 
the  fixed point is shown in Fig. \ref{fig:fixed point}(a).
The largest amount of energy remains in the center-of-mass degree of freedom until $t\approx 400$. 
For an energy density $h=0.01$, the average momentum per particle along these trajectories is $\sqrt{h}\sim 0.1$
implying that the center of mass keeps its energy for almost $80$ reflections with the wall.
Afterwards, the energy is distributed among the other normal modes within a few further reflections.
This observation makes it reasonable to introduce the notion of a relaxation time $\tau_r$, {\em i.e.} the time scale
on which the energy mode distribution starts to spread associated with a non-negligible width in the set of modes: 
\begin{equation}
    \tau_r= \min \left\{ t\left| E_{\text{com}}(t)< \frac{E_{\text{com}}(0)}{2}\right.\right\} \, .
\end{equation}
The results for $\tau_r$ are presented in Fig. \ref{fig:fixed point}(b,c) for different energy densities 
(averaged over $100$ different initial conditions), showing, on the whole, a moderate increase with increasing particle
number.

Our numerical analysis shows moreover that $\tau_r$ increases exponentially 
with the ratio $E_{\text{com}}(0)/E$ towards the fixed point.
This is a further indication, that the fixed point becomes unstable with increasing particle number. 

This trend is stabilized with increasing particle number $N$: we could not find any recurrences as in FPUT model, even when going to significantly longer times (of the order of $10^5$). Together with the exponential sensitivity of the relaxation time $\tau_r$ with respect to the initial condition, this lets us claim that the rigid-body fixed point stays unstable in the large $N$ regime.
Eventually it is worth noting that the above defined relaxation time scales as the period of the motion inside the box when varying $h$: rescaling by the period, {\em i.e. }multiplying by $\sqrt{h}$, our data show a fair collapse for the relaxation time.

\subsubsection{Localized solutions} \label{localquench}

The second family of regular initial conditions generalizes the asymmetric high energy solution that centered a stable island for $N=2$. For illustrative purposes we pick the maximally asymmetric such solution. This {amounts to dealing with the} lattice scale physics when looking at the system as a discretized field theory.
More precisely consider {the excitation of a single particle},  with the following initial condition:
\begin{align}
\begin{split}
x_i(0)&=0.5\,, \, 1\leq i\leq N \\
p_i(0)&=
 \begin{cases}
\sqrt{E} & \,i=1 \, , \\
0  & \, \text{otherwise} \, ,
\end{cases}\\ 
\end{split}\label{CI_quench}
\end{align}
{\em i.e.} i.e the energy is stored in the kinetic energy of the first particle. This initial condition is also relevant to investigate the analogy in larger dimension of the second stable fixed point visible for $N=2$ in Fig.\ref{fig:Poincared}(c). 
In order to test, whether the dynamics of the remaining particles is directly affected by the presence of the wall, we define the following quantity:
\begin{align}\label{maxx2}
  x_2^{\text{max}}=\max x_2(t)-0.5.
\end{align}
Whenever $x_2^{\text{max}}$ reaches the threshold value $0.5$, the particle next to the initially excited one is touching the wall. This stands for a test for energy relaxation: if the next particle is not excited enough to touch the wall, this means that energy equipartition among all particles cannot take place.\\ 
We start with $N=3$ particles. The equation of motion for the neighbor of the initially excited particle is given by, see~(\ref{EOM}):
\begin{align}
    \ddot{x}_2=4 (x_1-x_2)+4(x_3-x_2).
\end{align} 
By symmetry one has further $x_2=x_3$, and the equation of motion reduces to
\begin{align}
    \ddot{x}_2+\omega_0^2 x_2=\omega_0^2 x_1.\label{x2_driven}
\end{align}
where $\omega_0=2$ is the nonzero mode frequency of the chain with $3$ particles.
We now solve an approximating problem when the excitation energy $E$ is large. 
The dynamics of the initially excited particle, at the site $1$, is identified with a free particle, i.e. not feeling the interaction with its neighbors. Its neighbor is then treated as a driven harmonic oscillator following (\ref{x2_driven}).\\
The solution of the equations of motion inside a box of length $1$, obeying the initial conditions (\ref{CI_quench}), is simply given by a periodic triangular function:
\begin{align}
\begin{split}
x_1(t)&=
 \begin{cases}
2\sqrt{E}t-2 k +\frac{1}{2}, & 2k-\frac{1}{2}\leq 2\sqrt{E}\; t \leq 2k+\frac{1}{2} \, , \\
-2\sqrt{E}t+2 k +\frac{3}{2}, & 2k+\frac{1}{2}\leq 2\sqrt{E}\; t \leq 2k+\frac{3}{2} \, 
\end{cases}
\end{split}
\end{align}
with $k$ an integer number.
This can be rewritten as a Fourier series:
\begin{align}
    x_1(t)=\frac{1}{2}+\frac{4}{\pi^2}\sum_{k\ge 0} \frac{(-1)^k}{(2k+1)^2} \sin({(2k+1) \omega\, t}),\label{freepartbox}
\end{align}
with $\omega=2 \pi \sqrt{E}$ is the frequency of oscillations inside the box in our units.\\
The expression (\ref{freepartbox}) is then inserted as a driving for the nearest neighbor $x_2(t)$. This yields to the following form for the solution of (\ref{x2_driven}):
\begin{widetext}
\begin{align}
    x_2(t)=A \sin(\omega_0 t)+\frac{1}{2}+\frac{4}{\pi^2}\sum_{k \ge 0}  \frac{\omega_0^2}{\omega_0^2-(2k+1)^2 \omega^2}\frac{(-1)^k}{(2k+1)^2}\sin({(2k+1) \omega\, t}),
\end{align}
\end{widetext}
where $A$ is the amplitude of the homogeneous part. It can be determined by the initial conditions to be

\begin{align}\label{homo}
    A= -\frac{4}{\pi^2} \sum_{k \ge 0}  \frac{\omega_0\omega}{\omega_0^2-(2k+1)^2 \omega^2}\frac{(-1)^k}{(2k+1)}.
\end{align}

$x_2^{\text{max}}$ is then given by
\begin{align}
    x_2^{\text{max}}=|A|+B,
\end{align}
where $B$ is the amplitude of the driven term
\begin{align}\label{driven}
    B=\frac{4}{\pi^2}\sum_{k\ge 0}\frac{\omega_0^2}{\omega_0^2-(2k+1)^2 \omega^2}\frac{1}{(2k+1)^2}.
\end{align}
\begin{figure}[!ht]
  \centering
  \includegraphics[width=0.8\linewidth]{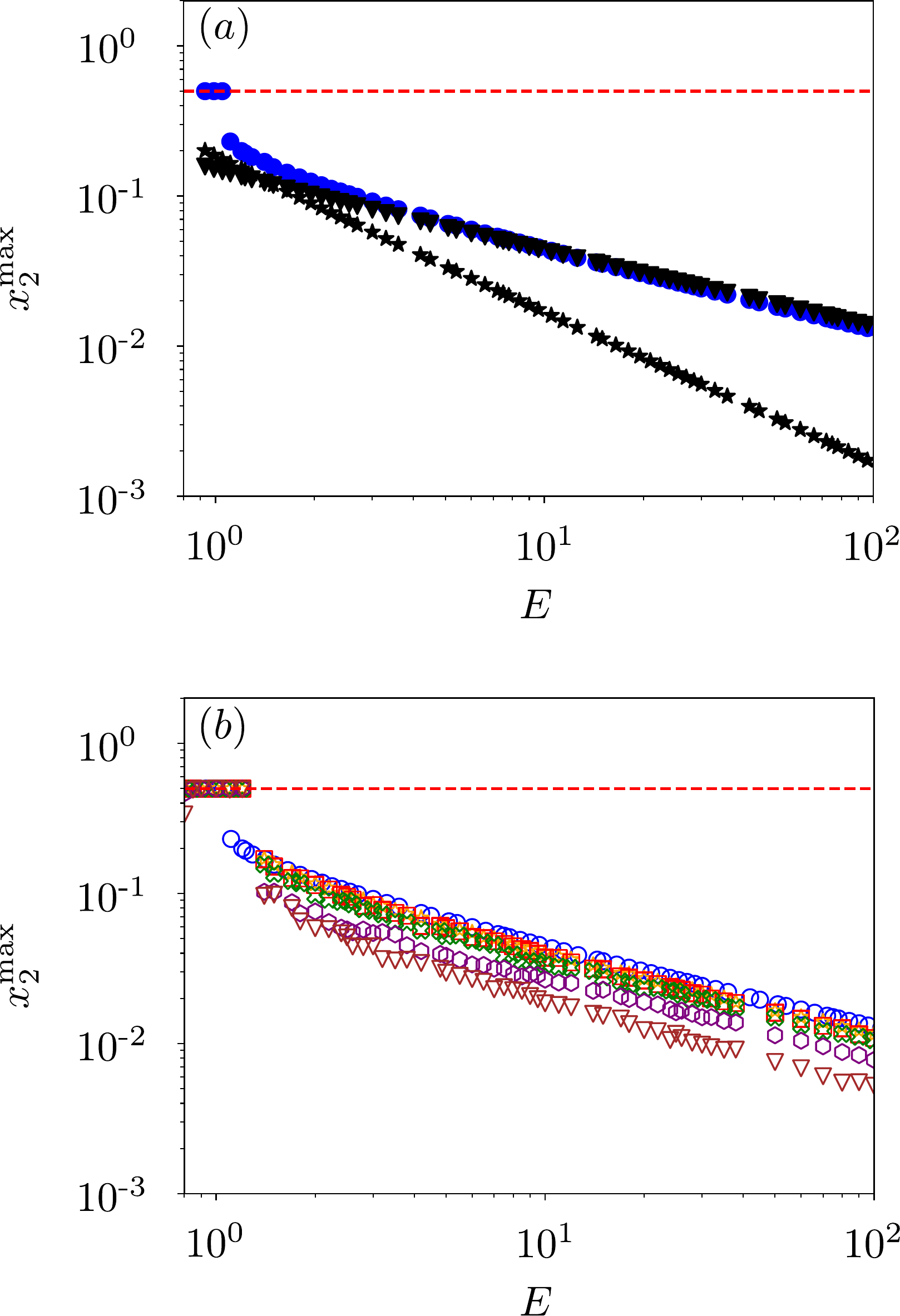}
  \caption{Amplitude of the motion of the nearest neighbor particle after {the excitation of a single site} as a function of the energy of the {excitation}. The dashed red lines indicate the threshold at which the neighbor particle touches the wall for a time up to $10^3$ . (a) $N=3$. Blue \new{circles}: numerical solution of the full chain. {Black triangles:} Homogeneous part of the analytical solution given by Eq.~(\ref{homo}). {Black stars:} Driven part of the analytical solution given by Eq.~(\ref{driven}). 
 (b)  Variation of $x_2^{\text{max}}$ as a function of the number of particles $N$. Blue circles: $N=3$. Orange stars: $N=5$. Red squares: $N=7$. Green crosses: $N=11$. Purple hexagons: $N=33$. Brown triangles: $N=65$.}
\label{fig:relax}
\end{figure}

The comparison between the numerical obtained amplitudes and the analytical approximations can be seen in Fig.~\ref{fig:relax}~(a). The separation between a fast moving free particle and its neighbors being driven by it, provides a very efficient approximation for large energies $E\gg 1$.
For energies around $E\simeq 1$, this approximation breaks down. This is clearly expected, since in that regime the dynamics of the initially excited particle becomes strongly affected by the interaction with its neighbors as the interaction energy is comparable to its kinetic energy. Therefore it does not follow the trajectory of a free particle as in Eq.~(\ref{freepartbox}). Moreover it is remarkable, that $x_2^{\text{max}}<0.5$ for large energies. This means that the dynamics of the neighbors of the highly excited particle is not affected by the presence of the wall, hence the energy sharing is strongly suppressed. \\
It is possible to build a similar simple approximation for larger particle numbers: $N>3$. Our numerical results in Fig.~\ref{fig:relax}~(b) show that this phenomenon crucially persists for larger values of $N$. Indeed, one can now describe the intuition behind this behavior: the central particle generates a high frequency drive acting on the rest of the chain. As long as the smallest frequency in this drive lies well above the bandwidth of the chain in the linear approximation, the response is strongly non-resonant and hence weak. Further, at these frequencies, the chain only supports evanescent waves and so the energy deposited into the central site cannot escape to infinity. This leads us to the conclusion that there is an absence of energy relaxation in the thermodynamic limit for the initial conditions of the type (\ref{CI_quench}). Perhaps we can extend these to non-zero energy density states by creating a super-lattice of such ``hot''sites but we have not investigated this carefully. We note that this is the analog of the KAM question in this system where we weakly couple the nonlinear degrees of freedom in each individual stadium. Absent the coupling, the system is integrable.

\subsubsection{Quasi-linear solutions} \label{noninteractingexcitations}

Next we discuss another family of low energy solutions which do not have any analog for $N=2$. These are sensitive to the non-linearity for small times but then become insensitive to it for a very long time, probably of the order of the Poincar\'e recurrence time for the linear problem which is clearly extremely long for large systems, see below.

Let us start with the simple observation that any initial condition, which leads to a time evolution where each position of the chain $x_i(t)$ obeys
$$ 0 < x_i(t) < 1, \ 1\le i \le N\ ,$$
is also an acceptable solution for the problems with the wall. Those solutions follow a linear time evolution, identical to the free chain. In particular those initial conditions have a Lyapunov spectrum which is trivial: every Lyapunov exponent is exactly $0$. The conserved quantities are the energy of the linear modes of the harmonic chain. Among those solutions there is a particularly interesting class: the solutions which start with a zero momentum of the center of mass. We found that, quite surprisingly, those solutions can be deformed in the presence of the walls to solutions, which will first touch the walls then 
follow a purely linear time evolution.

It may be fruitful at this stage to draw an analogy with the Caldeira-Leggett model \cite{caldeira1983quantum}, which has become one of the paradigmatic model for classical and quantum open systems. In that model a particle is in contact with a thermal bath, which leads to friction. In our model, for those initials conditions both initially excited particles experience a partial damping (some energy is leaking to the other site) followed by a long sequence of linear time evolution. Of course, for a finite chain, there will a Poincar\'e recurrence time, where the chain goes arbitrarily close to its initial configuration. In a thermodynamic perspective, {\em i.e.} when sending $N$ to infinity, this recurrence time diverges. Hence the time evolution, after a short damping episode, becomes linear and never feels the walls again. Due to this effectively integrable long time behavior, this set of initial conditions encodes a lack of thermalization despite some contacts with the walls. We found that those trajectories form a continuous family parameterized by their total energy $E$, which is bounded when required to observe this late linear evolution.
Still those initial conditions are not creating strictly stable regions in the phase space: any perturbation of such a trajectory, leading to a nonzero center-of-mass momentum, is likely to be ergodic.

\begin{figure}[!ht]
  \centering
  \includegraphics[width=0.8\linewidth]{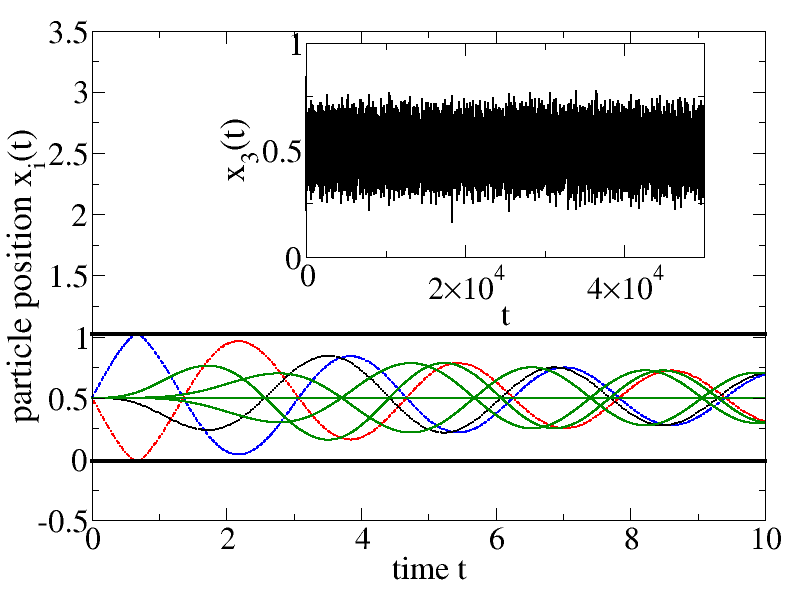}
  \caption{Trajectories of the particles inside a chain of length $N=64$. The initial conditions are given by (\ref{CI_quasilin}) with $E=10/\pi^2$, hence the rattling frequency is slightly off the band of the linear modes. Blue \new{dashed line}: $x_1(t)$. Red \new{dotted dashed line}: $x_2(t)$. Black \new{dotted line}: $x_3(t)$. Green \new{solid lines}: examples of $x_i(t)$ for $i\ge 4$. Inset: $x_3(t)$ on a longer time range.}
\label{traj_quasilin}
\end{figure}

To make the description clearer we choose $N=64$ and look at the following initial conditions:
\begin{align}
\begin{split}
x_i(0)&=0.5\,, \, 1\leq i\leq N \\
p_i(0)&=
 \begin{cases}
+\sqrt{E/2} & \,i=1 \, , \\
-\sqrt{E/2} & \,i=2 \, , \\
0  & \, \text{otherwise} \, ,
\end{cases}\\ 
\end{split}\label{CI_quasilin}
\end{align}
{\em i.e.}  every particle is at rest in the middle of the box, save two, which are given an initial velocity. For small enough $E$ this leads to solutions never reaching the box ends, hence trivial solutions of the problem with walls. In Fig.~\ref{traj_quasilin} it is shown that, for moderate value of $E$, the initially excited particles are touching the wall exactly once. They redistribute some energy to the chain. Their remaining energy is not enough to have them touching the wall a second time and the whole chain then follows a free evolution. In particular, one cannot see any contact with the wall on a time range, which is several orders of magnitude larger than both the traveling time inside the box (of order $1$), and the longest period of the linear mode (of order $N$).
Using the method described in \cite{zhang2017witnessing} we estimated the Poincar\'e recurrence time for the size $N=64$ to be of the order of $10^{32}$. More precisely this time was estimated to get a revival at a distance of the initial point less than $5\%$. This is significantly less than any closest approach distance seen in the Inset of Fig.~\ref{traj_quasilin}. This is the reason why we choose here the value $N=64$: we could not have an estimate of the recurrence time for larger values of $N$. Nevertheless we could run simulations for the chain for $N$ up to $N=1024$ (data not shown) and see the regime of linear evolution last over a time range longer than any other above mentioned time scales.

\subsection{About the continuum limit}

After reviewing some explicit examples of regular initial conditions, we discuss the continuum limit of the model in case, following the FPUT case, it sheds light on such matters.
This limit is achieved when one replaces the discrete particle positions $x_i(t)$ by a continuous scalar field $\phi(x,t)$.
First one may consider the presence of only one wall in the (scalar) field space at $\phi=0$. The wall can be seen as the limit of a smooth confining potential. One option for the potential is:
\begin{equation}
    V_1(\phi)=\frac{e^{-\alpha \phi}}{\alpha},\ \alpha>0\ .
\end{equation}
In the limit of $\alpha\to\infty$ the field $\phi$ is constrained to be non negative, hence a wall effect.
The Euler-Lagrange equation for the field theory with this potential are easy to obtain:
\begin{equation}
    \partial_t^2\phi -\partial_x^2 \phi = e^{-\alpha \phi} \label{EOMLivouille}\ ,
\end{equation}
where one recognizes the Liouville field theory, which is known to be integrable~\cite{PhysRevD.26.3517}.

Next one can repeat the same game for a field constrained in a one-dimensional box, say $0\le \phi \le 1$. The smoothing potential is now 
\begin{equation}
     V_2(\phi)=\frac{e^{-\alpha \phi}+e^{\alpha (\phi -\phi_0)}}{\alpha},\ \alpha>0 \label{smoothwalls2}
\end{equation}
where $\phi_0=1$ is the width of the box in the limit $\alpha\to\infty$. In that case we found this leads to a deformation of the sinh Gordon field theory, which is also integrable, see App.~\ref{mapSinhGordon}.

To summarize we believe that the underlying integrable continuum limit is deeply singular, the reflection on the wall leads to discontinuities in the time-derivative of the field, and a continuum approximation requires a significant effort to be justified. {Interestingly when devising a smoothed version of the model with a steep trapping potential instead of hard walls, this leads to a fully integrable field theory.} 

\section{Recovery of statistical mechanics for the $n=1$ model: Canonical averages} 
\label{recovery of statistical mechanics}

In the following we study the validity of the (classical) ergodic hypothesis for statistical properties such as the mean
energy per particle and spatial two-point correlator. We aim to compare the statistical average performed within the
canonical ensemble, and the time-averages along few very long trajectories using molecular dynamics simulations.
While the canonical averaging implicitly assumes global ergodicity, the existence of non-ergodic phase space regions leads to deviations in the molecular dynamics results.
As was shown in the previous section, {the excitation of a single site} may contribute to enhance two particle- correlations. The deviations between molecular dynamics and the canonical ensemble are therefore a useful tool to get a quantitative understanding for the size of these non-ergodic regions.

Since the Hamiltonian of our system contains only nearest-neighbor interactions, the corresponding partition function
can be computed based on a transfer matrix approach \cite{PhysRev.60.252,FENDLEY2002411,Bellucci2013}. 
Here we develop the main ideas and provide details of the transfer matrix approach for our problem 
in App.~\ref{Transfer matrix method}. 
Consider a partition function of the form
\begin{widetext}
\begin{equation}
  Z(\beta)=\left(\frac{\pi}{\beta}\right)^{N/2}\int_0^1 \,\ud x_1\dots\int_0^1 \,\ud x_N T_\beta(x_1,x_{2})\dots  T_\beta(x_{N-1},x_{N}) T_\beta(x_{N},x_{1})
  \label{defZbeta}
\end{equation}
\noindent
\end{widetext}
where the transfer operator $T_\beta$ is defined via a symmetric kernel $T_\beta(x,y)$ on a compact space. There is a discrete set of eigenvalues $\lambda_l(\beta)$ for the integral equation \cite{Tricomi}
\begin{equation}
    \int_0^1 T_\beta(x,y)f_l(y) \ud y =\lambda_l(\beta) f_l(x)\, , \label{eig_eq_Tl}
\end{equation}
{where $f_l$ is the eigenfunction associated to $\lambda_l(\beta)$}.
The partition function of a chain with $N$ particles, defined in (\ref{defZbeta}) can be rewritten with those eigenvalues
\begin{equation}
    Z(\beta)=\left(\frac{\pi}{\beta}\right)^{N/2} \text{Tr}(T_\beta^N)=\left(\frac{\pi}{\beta}\right)^{N/2}\ \sum_{l=0}^{+\infty} \lambda_l(\beta) ^N\ .
\end{equation}
Since the spectrum of $T_\beta$ is discrete for positive $\beta$, only the largest term $\lambda_0(\beta)^N$ is relevant 
in the limit $N\rightarrow \infty$.
The average energy per particle is given by \cite{FENDLEY2002411}
\begin{equation}\label{ekin}
    h(\beta)=-\frac{1}{N} \frac{\partial }{\partial \beta} \log(Z(\beta))
    \simeq \frac{1}{2\beta}-\frac{\lambda_0'(\beta)}{\lambda_0(\beta)} \, .
\end{equation}
For our model the kernel of the transfer operator $T_\beta$ is
\begin{equation}
    T_\beta(x,y)=\exp\left[-\beta (x-y)^2\right], \quad 0\leq x,y\leq 1\, .
\end{equation}

\begin{figure}[ht]
    \centering
	\includegraphics[width=\linewidth]{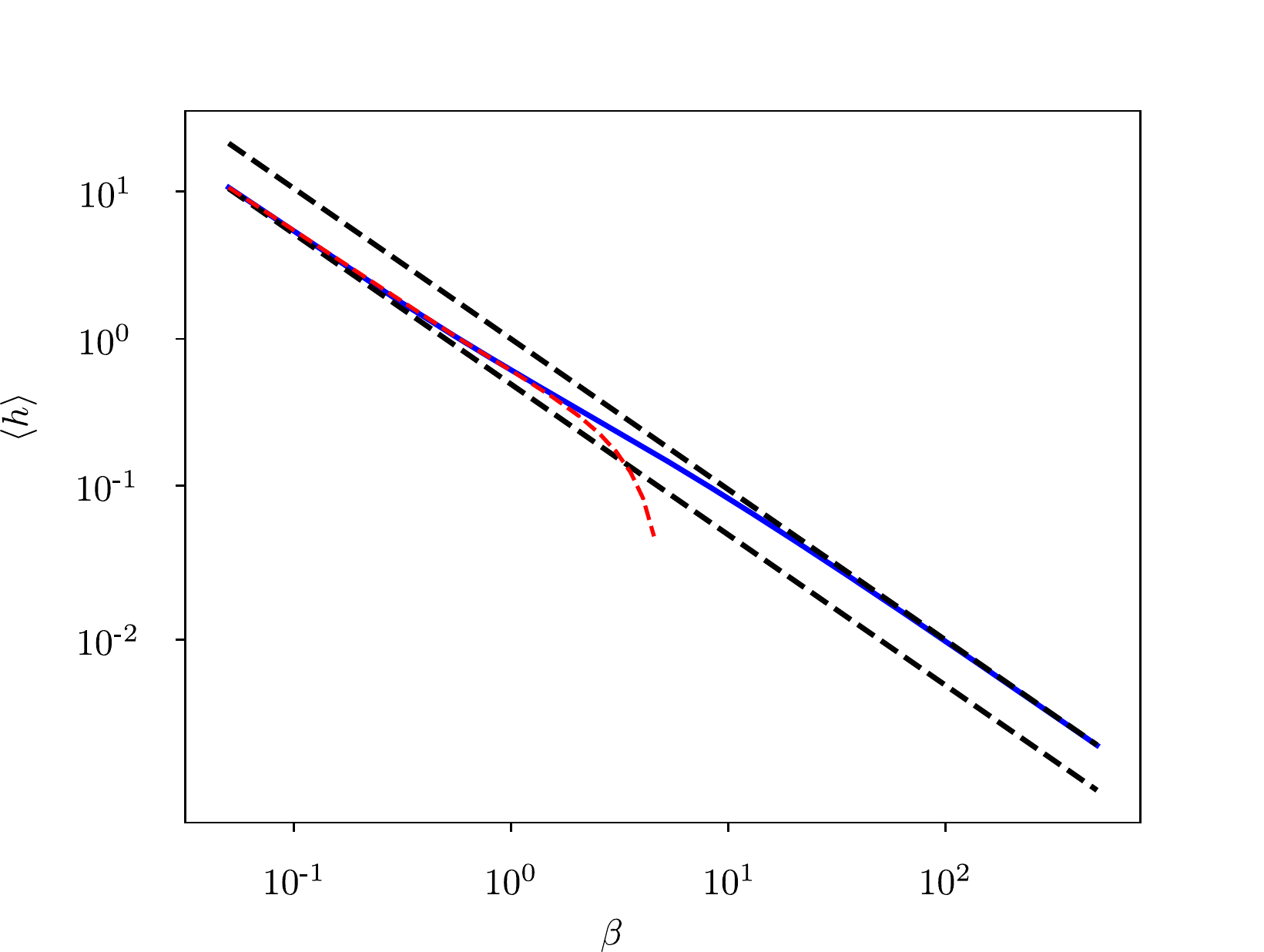}	
	\caption{Average energy density $h$ as a function of the inverse temperature $\beta$.
         Solid blue line: Canonical average energy density; 
         The black dashed lines indicate the trivial energy-temperature relations 
         $h(\beta)=\frac{1}{\beta}$ for two quadratic degrees of freedom per particle and 
         $h(\beta)=\frac{1}{2 \beta}$ for one degree of freedom per particle. 
         Those limiting cases are further justified in App.~\ref{Transfer matrix method}. 
         The red \new{dashed} curve shows the refined high-temperature approximation Eq.~(\ref{hbeta_highT}).}
	\label{fig:Temp}
\end{figure}

As we could not analytically solve the eigenvalue equation (\ref{eig_eq_Tl}), we discretized the space to
convert the integral equation into a linear system. Taking the matrix defining this system of size $10^3 \times 10^3$ was enough to ensure the numerical error to be less than $1\%$. 
The result for the temperature-energy relation is shown in Fig.~\ref{fig:Temp}.
The high- and low-energy limits of the model already indicated a non trivial temperature dependence of the energy density.
In view of our previous considerations in Sec. \ref{model} the system resembles, on the one hand, $N$ independent trapped particles 
in the high energy limit, implying one quadratic degree of freedom per particle in the limit $\beta \rightarrow 0$. 
On the other hand, for $\beta \rightarrow \infty$ one expects an energy-temperature relation similar to a 
harmonic oscillator and thus two quadratic degrees of freedom per particle. Our numerical canonical solution 
confirm that this is indeed the case. 
 The corresponding solid blue curve in Fig.~\ref{fig:Temp} shows a crossover from $h=0.1$ and $h=1$ between the regimes when there are approximately one or two quadratic degrees of freedom per particle.
The high-temperature behavior is further analytically supported and understood in terms of a more advanced approximation
for the leading eigenvalue of the transfer operator $T_\beta$, see App.~\ref{Transfer matrix method}. 

The transfer matrix approach can also give predictions for the spatial correlation functions, 
see App.~\ref{Transfer matrix method}. Consider the two-point correlator
\begin{equation}\label{correlation}
    C(i)=\langle x_{j} x_{j+i}\rangle-\langle x_{j}\rangle \langle x_{j+i}\rangle \, .
\end{equation}
\noindent
Without long-range order it is decaying exponentially as $C(i)\sim \exp(-i/\xi)$, where $\xi$ denotes the correlation length.
The dominant contribution of the correlation function is of order $(\frac{\lambda_1}{\lambda_0})^{i}$, where $\lambda_1$ is the second largest eigenvalue of $T_\beta$, as shown in
App.~\ref{Transfer matrix method}, thus leading to the correlation length
\begin{equation}\label{decay length}
    \xi(\beta)=\frac{1}{\log\left(\frac{\lambda_0(\beta)}{\lambda_1(\beta)}\right)} \, .
\end{equation}
Using a recently derived approximation \cite{bogomolny2019formation} for the integral equation (\ref{eig_eq_Tl}) one can also check that the critical exponent for the correlation length coincides with the value from the universality class of Ising model:
\begin{equation*}
  \xi \propto T^{-1}, \quad T\to 0\ .
\end{equation*}
\begin{figure}[ht]
  \includegraphics[width=\linewidth]{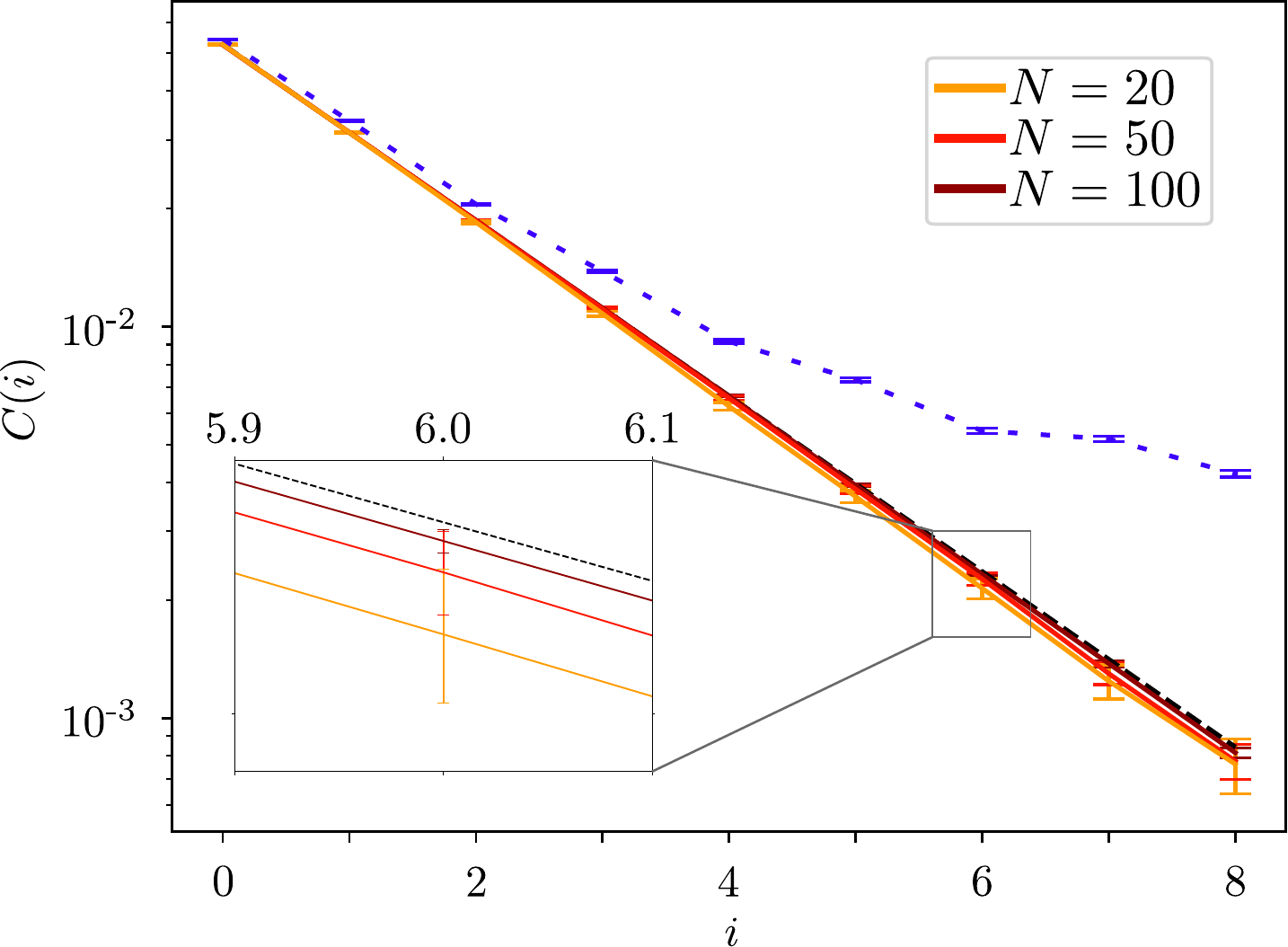}
  \centering
  \caption{
    \new{Two-particle correlation function $C(i)$, Eq.~(\ref{correlation}). Molecular dynamics simulations are performed for different particle numbers  for an energy density $h=0.1$ for scaled times up to $t=10^6$  and arbitrary initial conditions. The black dashed curve shows the results of the transfer matrix method with $\beta=8.55 $. The blue dotted curve shows $C(i)$ near to the ”rigid body” trajectory. Inset: zoom into $C(i)$. An increasing deviation between molecular dynamics and canonical ensemble with decreasing particle number is visible.}
  }
  \label{fig:Comparison}
\end{figure}

After uncovering the temperature energy relation, it is now possible to compare the predictions of the canonical ensemble 
average and molecular dynamics \cite{doi:10.1063/1.1730376}. To this end we considered the two-point correlation 
function $C(i)$ Eq.(\ref{correlation}) with $\langle x_{i}\rangle=0.5$ due to symmetry.
In our numerical results the average is taken over the site index $n$, and the canonical ensemble or single trajectory until $t=10^6$ to bound the absolute error by $2 \cdot 10^{-4}$. 
Quasi-integrable dynamics in or close to the identified stable island in \ref{stability} gives rise to large correlations 
of neighboring particles. Since the canonical ensemble averages over the entire phase space, one expects the molecular 
dynamics to give smaller predictions for the correlation function.
As can be seen in Fig.(\ref{fig:Comparison}), this is indeed the case. 
The difference between canonical ensemble and molecular dynamics increases with decreasing particle number $N$. 
This shows the increasing impact of non-ergodic phase space regions to the global phase space dynamics of the system when the total 
number of particle is reduced. On the opposite, \emph{i.e.} when increasing $N$, the data indicate better and better agreement between both procedures.

\section{Two component ($n=2$) scalar model}\label{stadium_results}

So far we have been discussing a single scalar field attached to each lattice site. Adding a second scalar field significantly enriches the model as then the local dynamics at each site can be made chaotic. Even more, one could devise in advance which type of chaotic dynamics (weakly or strongly mixing) each sites will follow.

First one may ask for each site being trapped in a rectangular billiard. In that case both the Hamiltonian (\ref{bft_vect}) and the boundary conditions separate between the $x$, and $y$ directions. Therefore all our previous discussion about the $n=1$ can be immediately transcribed here. One could for example devise initial conditions which lead to quasi-linear evolution for the $x$ component of the field at each site, whereas the $y$ component follows a localized behavior.

The picture changes drastically, when the boundary conditions start to couple both components of the local field. In particular we consider the Hamiltonian (\ref{bft_vect})  where each local field $x_i(t),y_i(t)$
forms a two-dimensional vector, whose endpoint is confined inside the stadium billiard so that the local dynamics is now chaotic.

\begin{figure}[!ht]
  \centering
  \includegraphics[width=0.8\linewidth]{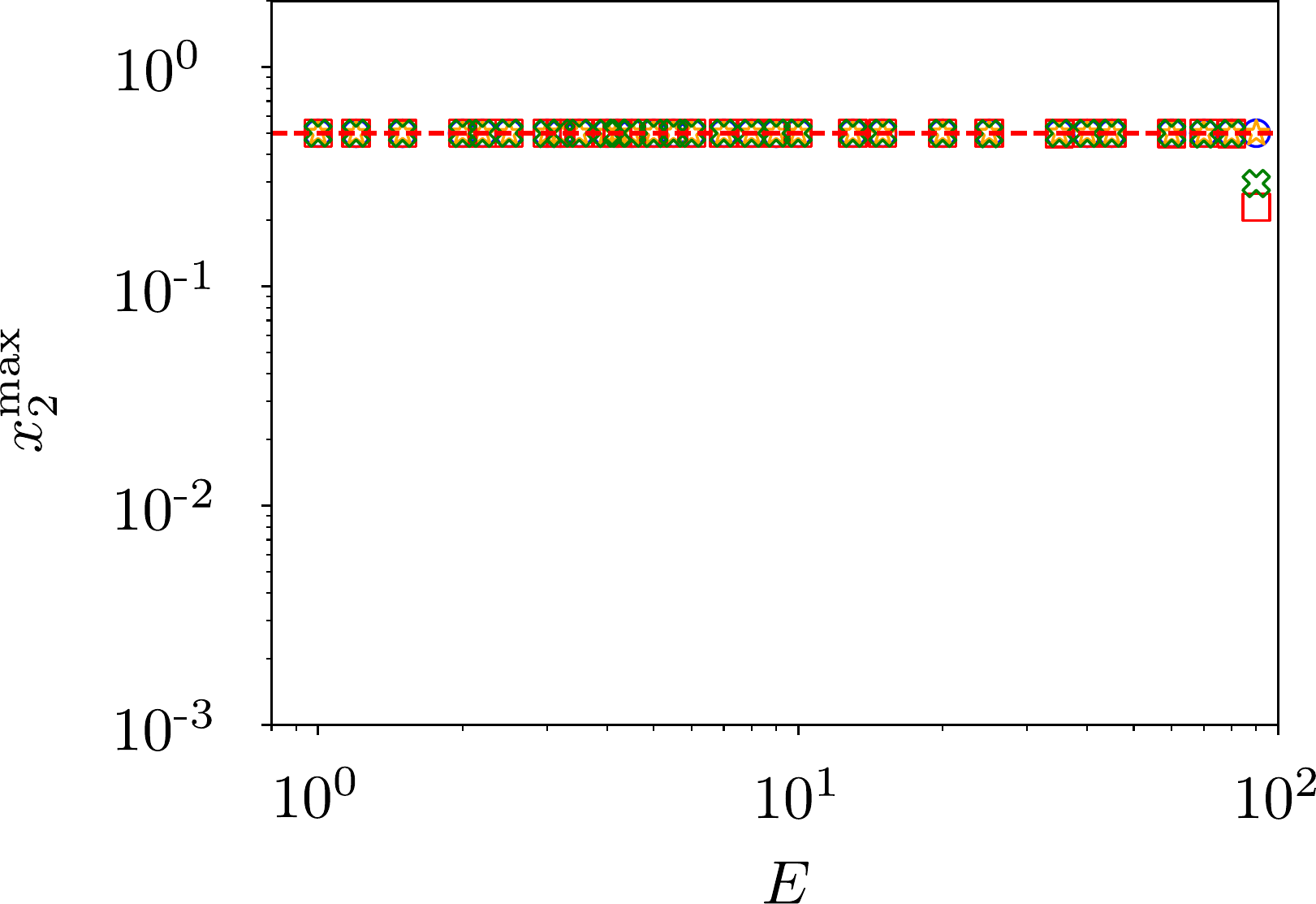}
  \caption{(a) Same as Fig.~\ref{fig:relax}~(b) when each site of the chain carries a vector, whose amplitude is now confined inside a stadium billiard. Blue circles: $N=3$. Orange stars: $N=5$. Red squares: $N=7$. Green crosses: $N=11$. Purple hexagons: $N=33$. Brown triangles: $N=65$.}
\label{fig:billiardrelax1}
\end{figure}

The \old{quench} protocol \new{of a single site excitation} to search for quasi-localized solutions is now as follows: one particle is given a very large initial kinetic energy in an arbitrary direction, whereas the others stand still.
We repeated the same analysis as above, and computed the maximum amplitude of $y_2$. The results, which are shown~in Fig.~\ref{fig:billiardrelax1}, clearly show that the energy sharing is no longer suppressed (in comparison with Fig.~\ref{fig:relax}). Instead one can see that the initial driving of one particle excites its nearest neighboring particles in such a way that they will be affected by the presence of the wall.\\

\begin{figure}[!ht]
  \centering
  \includegraphics[width=0.8\linewidth]{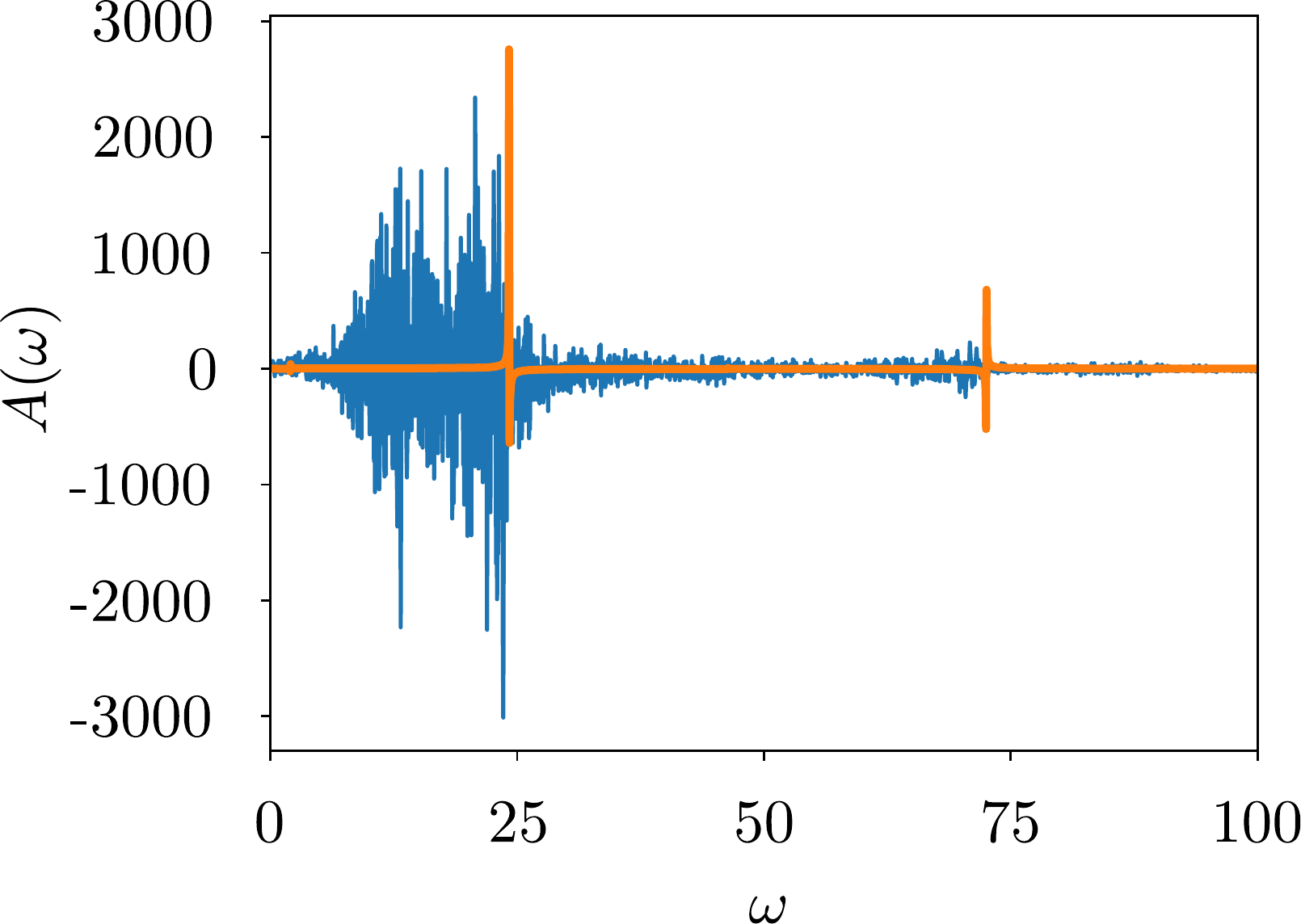}
  \caption{Fourier spectrum for the motion of the first,excited particle, with $E=15.0$ and $N=3$. Blue/Dark grey: in the stadium billiard ($n=2$). Orange/Light grey: in the box ($n=1$).}
\label{fig:billiardrelax2}
\end{figure}

This can be understood considering the Fourier spectrum of the driving particle motion, as can seen in Fig.~\ref{fig:billiardrelax2}. We considered the $y$-component of the driving particle motion up to times of $T=1.3 \cdot 10^2$, {\em i.e.} before achieving energy relaxation. Then a Fourier transform was performed on $y_1(t)$. While the Fourier spectrum of the box is discrete and far off-resonance, the chaotic motion in the stadium billiard leads to a continuous spectrum. The low-frequency components are closer to resonance and allow now for energy sharing. This contribution originate\new{s} from time intervals where the particle is propagating mostly in the $x$-direction. The low $y$-component leads to an enhanced energy sharing.

\section{Conclusions}

We have introduced a family of models of coupled classical nonlinear oscillators. These models can live on $d$ dimensional lattices and involve $n$ scalar fields per site which are confined to a chosen domain (the billiard table or stadium). We focused on the simplest situation of a chain ($d=1$) of $N$ sites, where the scalar field ($n=1$) at each site is confined inside a box. We performed extensive numerical simulations for size up to $N=30$, and find that the system is ergodic for randomly chosen initial conditions. More precisely, the long time limit of the two-point correlator agrees well with the predictions of statistical mechanics. Unlike another famous nonlinear chain, that of FPUT, we did not see any evidence of recurrences suggestive of \old{integrability}\new{an integrable continuum limit}. This is a little surprising, as the continuum limit of our system is itself a particular limit of the integrable sinh Gordon system. However, it is likely that the integrability of the latter system is lost in the passage to the limit.

\old{While overall, we suggest our model obeys statistical mechanics in the large $N$ limit, we have also provided explicit families of initial conditions, which lead to non-ergodic behavior and absence of thermalization.}
\new{It was proven that a 'generic' trajectory, {\em i.e.} with random initial conditions is ergodic in the large $N$ limit hence statistical mechanics is applicable in that sense. But it is worth stressing that we also provided explicit families of initial conditions, which lead to non-ergodic behavior and absence of thermalization.}
Two were inferred from the $N=2$ case: one corresponds to the field being identical at every sites. This looks like a particular set for the initial data in the field theory obtained in the continuum limit. The other initial condition looks at the opposite limit with short wavelength (of the size of the mesh) fluctuations: this \new{"}local quench\new{"} type of initial conditions leads to a localized dynamics where the energy only leaks for a short period of time from the excited site to its neighbors. Remarkably we also identified a last continuous family of initial conditions where the chain starts to feel the non-linearities due to the wall, then follows the behavior of a linear harmonic chain. In future work it would be interesting to see if one can quantify the scaling with $N$ of the phase space volume for such solutions and whether there is a KAM approach for small $N$ about the decoupled well limit.

The natural next task is to study the model with $n=2$ that was introduced in this paper. We gave evidence that the chaos already present at the level of a single site destroys the localized solutions that we found---so the model is now considerably more chaotic. A more careful study of relatively small values of $N$ could be rewarding in that we hope to find that this model exhibits much stronger chaos and shows convincing evidence of effective ergodicity as it heads towards the infinite volume limit. Separately, it would be interesting to introduce disorder in our models to see if we can generate many-body localization on reasonably long time scales.

\begin{acknowledgments}
 RD thanks Matteo Sommacal for helpful discussions and providing Ref.~\cite{Tricomi} and M. Y. Lashkevich for discussing the connection with sinh Gordon. SLS would like to thank David Campbell for sharing his wisdom on chaotic matters.
RD, JDU and KR acknowledge funding from the Deutsche Forschungsgemeinschaft (DFG) through Project Ri681/14-1.
DH thanks Shivaji Sondhi for the kind hospitality during his stay at Princeton and the Max-Weber-Program of 
the Bavarian Elite Program for supporting this stay financially.  SLS acknowledges support from the United States Department  of  Energy  via  grant  No.   DE-SC0016244. Additional support was provided by the Gordon and Betty Moore Foundation through Grant GBMF8685 towards the Princeton theory program.
\end{acknowledgments}

\appendix
\section{Numerical Integration}
\begin{figure}[ht]
  \includegraphics[width=1.0\linewidth]{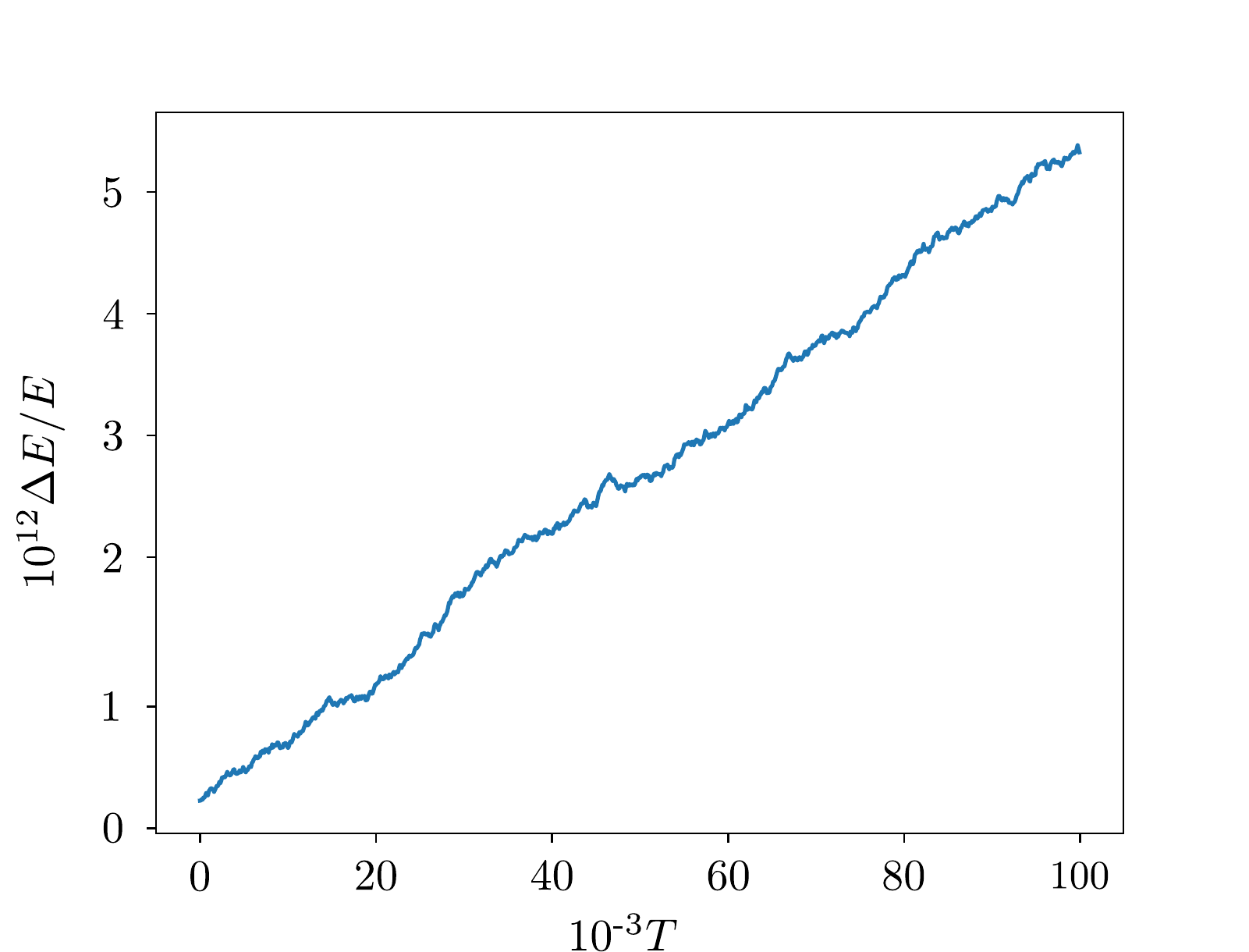}
  \centering
  \caption{Relative error $\frac{\Delta E}{E}$ for a typical trajectory. The starting energy density was $h=0.1$. With our algorithm we obtained a relative energy error $\frac{\Delta E}{E} \sim 10^{-11}$.} 
  \label{fig:energy}
\end{figure}
The time integrations were performed by an adaptive Runge-Kutta-algorithm of fourth order.
The default step size was $3 \cdot 10^{-4}$. After each step, the particle are tested, whether \new{each of them is still located inside its own box}. If it is not the case, the original coordinate is maintained and the step size is reduced by a factor $10$. This procedure is repeated, until a final step size less than $10^{-12}$ is reached. Finally, the sign of the momentum of the particle at the boundary is reversed and the step size set to its default value again \new{({\em i.e.} $3 \cdot 10^{-4}$)}. With that algorithm, we obtained a relative energy error of $\frac{\Delta E}{E} \leq 10^{-11}$ \new{for a total time of $10^5$}.
\section{Calculation of Lyapunov Exponents}
The calculation scheme for Lyapunov exponents is based on the methods described in \cite{PhysRevE.53.1485,doi:10.1063/1.5099446}.
For given equations of motions
\begin{equation}
   \mathbf{\dot{\Gamma}}(t)=\mathbf{F}(\mathbf{\Gamma}(t)),
\end{equation}
an infinitesimal deviation evolves according to
\begin{equation}
    \dot{\delta\mathbf{\Gamma}}(t)= \frac{\partial \mathbf{F}(\mathbf{\Gamma)}}{\partial \mathbf{(\Gamma)}}\Biggr|_{\mathbf{\Gamma}(t)} \delta \mathbf{\Gamma}(t),
\end{equation}
\noindent
In the case of coupled harmonic oscillators $\mathbf{F}(\mathbf{\Gamma})= \mathbf{F}\mathbf{\Gamma}$ is linear, therefore $\delta\mathbf{\Gamma} (t)$ can be calculated by a simple matrix exponential between two collisions:
\begin{equation}
    \dot{\delta\mathbf{\Gamma}} (t)= e^{\mathbf{F} t} \delta \mathbf{\Gamma}(0),
\end{equation}
\noindent
At times $t_j(\mathbf{\Gamma})$, one of the particle is reflected on the wall.
This can be described by the mapping
\begin{equation}
    \mathbf{\Gamma}'=\mathbf{M}\mathbf{\Gamma},
\end{equation}
\noindent
where $\mathbf{M}$ switches the sign of the momentum of the reflected particle.
According to \cite{PhysRevE.53.1485}, the mapping for the deviation problem is for the linear problem given by
\begin{equation}
    \delta \mathbf{\Gamma}'=\mathbf{M} \delta \mathbf{\Gamma}+ [\mathbf{M},\mathbf{F}] \mathbf{\Gamma} \delta \tau_c,
\end{equation}
\noindent
with  $\delta \tau_c=t_j(\mathbf{\Gamma}+\delta\mathbf{\Gamma})-t_j(\mathbf{\Gamma})$ the collision delay time for the deviated trajectory.\\
Repeating the steps in \cite{doi:10.1063/1.5099446}, after a reflection of the $i$-th particle the new deviations $\delta x_i'$ and $\delta p_i'$ can be expressed by the deviations $\delta x_i$ and $\delta p_i$ before the collision with the wall:
\begin{align}
\begin{split}
&\delta x_i' = -\delta x_i \\
&\delta p_i' = - \delta p_i- 4  \frac{(x_{i+1}-2 x_i+x_{i-1})}{p_i} \delta x_i 
\end{split}
\end{align}
\noindent
Here $x_i$, $p_i$ denote the coordinate and momentum of the $i$-th particle before the reflection, $\delta x_i$ and $\delta p_i$ the corresponding deviations. The other entries remain unchanged.\\ \\
In order to calculate the entire Lyapunov spectrum, we used the algorithm proposed by \cite{Benettin1980}.
\noindent
As a numerical check, the largest Lyapunov exponent was independently calculated by the algorithm presented in \cite{PhysRevA.14.2338} for a few random initial conditions. Both techniques gave the same result.

\section{The transfer matrix method}\label{Transfer matrix method}
This Section is based on \cite{Goldenfeld:1992qy}. It is here adapted to our present model.
The Hamiltonian of our model is
\begin{equation}
  H(\{ p_i,x_i\})=\sum_{i=1}^N p_i^2 + (x_i-x_{i+1})^2 + V(x_i)\ ,
\end{equation}
where the potential $V(x_i)$ stands for the confinement in a box for each particle: $0\le x_i\le 1$. It is also assumed that there are periodic boundary conditions $x_{N+1}=x_1$.
The canonical partition function for a given inverse temperature $\beta$ is (we choose units such that Planck's constant $h$ is unity):
\begin{equation}
  Z(\beta)=\int_{-\infty}^\infty \ud p_1\int_0^1\ud x_1\dots \int_{-\infty}^\infty \ud p_N\int_0^1 \ud x_N e^{-\beta H(\{ p_i,x_i\})}
\end{equation}
As usual the integration over the momenta is straightforward so there remains the multidimensional integral over the positions
\begin{equation}
  Z(\beta)=\left(\frac{\pi}{\beta}\right)^{N/2}\int_0^1\ud x_1\dots\int_0^1 \ud x_Ne^{-\beta \sum_i (x_i-x_{i+1})^2} \ . \label{Zdef}
\end{equation}
At this stage it is customary to introduce the following differential operator
\begin{eqnarray*}
  T_\beta: \mathrm{L}^2([0:1])&\longrightarrow&\mathrm{L}^2([0:1])\\
  f&\longmapsto& g\ ,
\end{eqnarray*}
with the defining formula
\begin{equation}
  g(x)\equiv(T_\beta f)(x)=\int_0^1 e^{-\beta (x-y)^2} f(y)\ud y
\end{equation}
As the kernel is smooth, and summable on the domain $(x,y)\in [0:1]\times[0:1]$, $T_\beta$ is a compact self-adjoint operator. Following the Hilbert Schmidt theorem, see e.g. \cite{Tricomi} p.~110, its spectrum is real, discrete and $T_\beta$  admits a spectral decomposition using its eigenvalues and corresponding eigenfunctions. Those are defined through the following equation:
\begin{equation}
  \int_0^1 T_\beta(x,y)  f(y)\ud y=\lambda f(x)\ , T_\beta(x,y)=e^{-\beta (x-y)^2}\ .\label{eigTbeta}
\end{equation}
Note that due to the trivial bound (for positive $\beta$):
$$ |T_\beta(x,y)|\le 1 $$
the eigenvalues are also bounded from above. Last one can show that $T_\beta$ is positive definite. Introduce $h(z)$ such that $T_\beta(x,y)=h(x-y)$. Here one has $h(z)=e^{-\beta z^2}$. Use that its Fourier transform is positive on the real axis:
$$ [\mathcal{F}h](k)\equiv \int_{-\infty}^\infty e^{-\ic k z}h(z)\ud z= \sqrt{\frac{\pi}{\beta}}e^{-k^2/4\beta}> 0\ . $$
Then for any function $f(x)$ in L$^2$, one has:
\begin{eqnarray*}
  \left<f,T_\beta f\right>&=&\int_0^1\ud x \int_0^1 \ud y e^{-\beta(x-y)^2}f(x)f(y)\\
  &=& \int_{-\infty}^\infty \frac{\ud k}{2\pi} \left| \int_0^1 f(x) e^{\ic k x}\ud x\right|^2 [\mathcal{F}h](k)\ , 
\end{eqnarray*}
and this quantity is zero iff $f(x)$ is identically $0$. This means that every eigenvalue is not degenerate and positive.

Eventually the eigenvalues of the transfer operator $T_\beta$ can be sorted in decreasing order:
$$ 1 \ge \lambda_0(\beta) > \lambda_1(\beta)> \dots > \lambda_{N-1}(\beta)> \dots>0\ .$$
The reason for introducing such an operator is because the partition function (\ref{Zdef}) can be rewritten as:
\begin{equation*}
  Z(\beta)=\left(\frac{\pi}{\beta}\right)^{N/2} \text{Tr}(T_\beta^N)
\end{equation*}
All thermodynamic quantities can be therefore expressed by the eigenvalues and eigenvectors of the transfer operator $T_\beta$. 
The transfer matrix approach can be also used to calculate expectation values or two-point correlation functions in space. Consider for example the expectation value $\left< x_i\right>$. It can be written in the form
\begin{widetext}
  \begin{equation}
    \hspace{-1cm}\langle x_{j}\rangle=\frac{1}{Z}\int_0^1\ud x_1\dots \int_0^1\ud x_N T_\beta(x_1,x_2)\dots T_\beta(x_{j-1},x_j)\, x_j \,T_\beta(x_{j} x_{j+1})\dots
    T_\beta(x_{N-1},x_N)T_\beta(x_{N},x_1)
    =\frac{1}{Z}\sum_{l=0}^\infty \lambda_l(\beta)^N \int_0^1 x |f_l(x)|^2\ud x\ , 
  \end{equation}
\end{widetext}
where $f_l$ stands for the normalized eigenfunction of $T_\beta$ associated to $\lambda_l(\beta)$ following (\ref{eigTbeta}).
Similarly the space correlation function can be expressed as
\begin{widetext}
\begin{eqnarray*}
  \hspace{-1cm}\langle x_{j}x_{j+i}\rangle&=&\frac{1}{Z}\int_0^1\ud x_1\dots \int_0^1\ud x_N T_\beta(x_1,x_2)\dots T_\beta(x_{j-1},x_j)\, x_j \,T_\beta(x_{j} x_{j+1}) \dots T_\beta(x_{j+i-1},x_{j+i})\, x_{j+i} \,T_\beta(x_{j+i} x_{j+i+1})\dots T_\beta(x_{N},x_1)\\ 
    &=&\frac{1}{Z}\sum_{l,m \ge 0} \lambda_l(\beta)^{N-i}\lambda_m(\beta)^{i} \left(\int_0^1 x f_l(x) f_m(x)\ud x\right)^2\ . 
\end{eqnarray*}
\end{widetext}
When going to the continuum limit $N\to\infty$, all the above formulas become significantly simpler. The partition function is well approximated by:
\begin{equation*}
   Z(\beta)\simeq\left(\frac{\pi}{\beta}\right)^{N/2} \lambda_0(\beta)^N,\ N\to\infty\ .
\end{equation*}
This approximation is very useful to compute the temperature-energy relation. The mean energy per particle is given by
\begin{equation}
  \label{energytemperature}
h(\beta)=- \frac{1}{N}\frac{\partial }{\partial \beta} \log(Z(\beta))\simeq \frac{1}{2\beta}-\frac{\lambda_0'(\beta)}{\lambda_0(\beta)},\ N\to \infty
\end{equation}
In the last equation, the contributions of $\lambda_i^N$ for $i\ge 1$, has been neglected as they are exponentially smaller in the large $N$ regime.
Similarly the two-point correlation function simplifies in this regime to
\begin{equation}
  \langle x_{i}x_{i+k}\rangle-\langle x_{i}\rangle\langle x_{k}\rangle\simeq \left(\int_0^1 x f_0(x) f_1(x)\ud x\right)^2 \left(\frac{ \lambda_1(\beta)}{ \lambda_0(\beta)}\right)^k
\end{equation}
In particular it varies with $k$ like $\exp[-k/\xi(\beta)]$, where the correlation length $\xi(\beta)$ is given by
\begin{equation}
  \xi(\beta)=\frac{1}{\ln\left(\frac{ \lambda_0(\beta)}{ \lambda_1(\beta)}\right)}
\end{equation}

\subsection{High temperature ($\beta\to 0$) regime}

At large temperature, or small $\beta$, the integral equation defining the transfer operator becomes very simple. Starting from the Taylor expansion valid for $\beta$ going to $0$:
\begin{equation*}
  e^{-\beta(x-y)^2}\simeq1-\beta (x-y)^2+\frac{\beta^2}{2}(x-y)^4 \ ,
\end{equation*}
the integral equation (\ref{eigTbeta}) becomes
\begin{equation}
  \int_0^1 \left[1-\beta(x-y)^2+\frac{\beta^2}{2}(x-y)^4\right] \phi(y)\ud y=\lambda \phi(x)\ .\label{eigTbeta0}
\end{equation}
Looking at the left hand side one can see that $\phi(x)$ is a fourth degree polynomial. Therefore one can put the trial formula
$$\phi(x)=a_4 x^4+a_3 x^3+a_2 x^2+a_1 x+a_0$$
into (\ref{eigTbeta0}) to solve the eigenvalue problem. In this regime, this becomes a linear system. More precisely putting a fourth degree polynomial into the integral equation leads to the following matrix eigenvalue equation:
\begin{widetext}
  \begin{equation}
    \left(\begin{array}{ccccc}
            \frac{\beta^2}{10} & \frac{\beta^2}{8} & \frac{\beta^2}{6} & \frac{\beta^2}{4} & \frac{\beta^2}{2}\\
            -\frac{\beta^2}{3} & -\frac{2\beta^2}{5} &-\frac{\beta^2}{2} & -\frac{2\beta^2}{3} & -\beta^2 \\
            -\frac{\beta}{5}+\frac{3\beta^2}{7} & -\frac{\beta}{4}+\frac{\beta^2}{2} &-\frac{\beta}{3}+\frac{3\beta^2}{5} &-\frac{\beta}{2}+\frac{3\beta^2}{4} &-\beta+\beta^2 \\
\frac{\beta}{3}-\frac{\beta^2}{4} & \frac{2\beta}{5}-\frac{2\beta^2}{7} & \frac{\beta}{2}-\frac{\beta^2}{3} & \frac{2\beta}{3}-\frac{2\beta^2}{5} & \beta-\frac{\beta^2}{2} \\ 
            \frac{1}{5}-\frac{\beta}{7}+\frac{\beta^2}{8} & \frac{1}{4}-\frac{\beta}{6}+\frac{\beta^2}{16} & \frac{1}{3}-\frac{\beta}{5}+\frac{\beta^2}{14} &                 \frac{1}{2}-\frac{\beta}{4}+\frac{\beta^2}{12} &  1-\frac{\beta}{3}+\frac{\beta^2}{10}
          \end{array}\right)
    \left(\begin{array}{c}
            a_4 \\ a_3 \\a_2 \\a_1 \\ a_0
          \end{array}\right)
  =\lambda
  \left(\begin{array}{c}
          a_4 \\ a_3 \\a_2 \\a_1 \\ a_0
        \end{array}\right)
\end{equation}
\end{widetext}
Although we cannot write an explicit expression for the largest eigenvalue in general, we can determine its Taylor series for small $\beta$:
\begin{equation}
  \lambda_0(\beta)\simeq 1-\frac{\beta}{6}+\frac{7\beta^2}{180},\quad \beta\to 0\ ,
\end{equation}
so that the partition function for the chain in the continuum limit $N\to\infty$, and in the regime of large temperature ($\beta\to 0$) is:
\begin{equation}
  Z(\beta)\simeq\left(\frac{\pi}{\beta}\right)^{N/2} \lambda_0(\beta)^N\simeq \left(\frac{\pi}{\beta}\right)^{N/2}\left(1-\frac{\beta}{6}+\frac{7\beta^2}{180}\right)^N\ .
\end{equation}
Following (\ref{energytemperature}) the equipartition theorem is:
\begin{equation}
  h(\beta)\simeq \frac{1}{2\beta}+\frac{1}{6}-\frac{\beta}{20},\ \beta\to 0\ .\label{hbeta_highT}
\end{equation}
Note that the constant term can also be recovered using first order perturbation theory in the coupling constant $g$.
\subsection{Low temperature ($\beta\to \infty$) regime}

The regime of small temperature, or large $\beta$, is the most interesting one. It is investigated using the trace of the resolvent of $T_\beta$, see e.g. \cite{Tricomi}. The eigenvalues of the transfer operator $T_\beta$ are the zeroes of a characteristic function $F(\lambda)$, which is analytic in the domain $\lambda\neq 0$. This function $F(\lambda)$ has also an exact converging expansion in the domain $\lambda >1$ using the properties of the kernel. 
Using approximating formulas for the kernel in the regime $\beta\to \infty$ we will derive an approximation for this expansion. Assuming analytic continuation one may obtain some information about the leading eigenvalues.

Start with the exact identity, see Eq.(18) p.72 in \cite{Tricomi},
\begin{equation}
  \frac{F'(\lambda)}{F(\lambda)}=\frac{1}{\lambda}\sum_{n=1}^\infty \frac{1}{\lambda^n} \textrm{Tr} K_n\ , \label{charF}
\end{equation}
where $K_n$ are the iterated kernels:
\begin{eqnarray*}
  K_1(x,y)&=&T_\beta(x,y)\\
  K_{n+1}(x,y)&=&\int_0^1 K_n(x,z)T_\beta(z,y)\ud z,\quad n\ge 1\, .
\end{eqnarray*}
The main remark is that those kernels can be easily estimated for large $\beta$. More precisely we will show by recursion that
\begin{equation}
  K_n(x,y)\simeq \left(\sqrt{\frac{\pi}{\beta}}\right)^{n-1}\frac{e^{-\beta\frac{(x-y)^2}{n}}}{\sqrt{n}},\quad n\ge 1\, .
\end{equation}
This assumption is trivially true for $n=1$. If it is assumed for $n$, then
\begin{eqnarray*}
  K_{n+1}(x,y)&=&\int_0^1 K_n(x,z)T_\beta(z,y)\ud z\\
 &\simeq& \left(\sqrt{\frac{\pi}{\beta}}\right)^{n-1}\frac{1}{\sqrt{n}}\int_0^1e^{-\beta\frac{(x-z)^2}{n}-\beta(z-y)^2}\ud z
\end{eqnarray*}
Next one uses that in the regime of $\beta\to\infty$ the integral can be approximated using a saddle point approach: the main contribution comes from the neighborhood of the minimum of
$$ g_{n+1}(z)=\frac{(x-z)^2}{n}+(z-y)^2\ .$$
This minimum is reached for $z=(x+ny)/(n+1)$, which is always in the prescribed range $[0;1]$ for every $n\ge 1$. Therefore one can extend the integration range to the whole real axis, and using that
$$ g_{n+1}\left(\frac{x+ny}{n+1}\right)=\frac{(x-y)^2}{n+1}\ ,$$
one gets
\begin{eqnarray*}
  \int_0^1 e^{-\beta\frac{(x-z)^2}{n}-\beta(z-y)^2}\ud z &\simeq& e^{-\beta \frac{(x-y)^2}{n+1}}\int_{-\infty}^\infty
 e^{-\beta \frac{n+1}{n}[z-\frac{x+ny}{n+1}]^2} \ud z\\
 &=&\sqrt{\frac{\pi}{\beta}\frac{n}{n+1}}e^{-\beta \frac{(x-y)^2}{n+1}}\, .
\end{eqnarray*}
Inserting this result in the definition of $K_{n+1}$ one gets
$$ K_{n+1}(x,y)\simeq\left(\sqrt{\frac{\pi}{\beta}}\right)^{n}\frac{e^{-\beta\frac{(x-y)^2}{n+1}}}{\sqrt{n+1}}\ ,$$
which ends the recursion proof.

Those approximations for each iterated kernels enables one to estimate the traces:
\begin{equation}
  \textrm{Tr} K_n\equiv \int_0^1 K_n(x,x)\ud x\simeq \left(\sqrt{\frac{\pi}{\beta}}\right)^{n-1}\frac{1}{\sqrt{n}}
\end{equation}
Then the right hand side of (\ref{charF}) can be rewritten
\begin{equation}
  \frac{F'(\lambda)}{F(\lambda)}\simeq\frac{1}{\lambda}\sqrt{\frac{\beta}{\pi}}\sum_{n=1}^\infty\frac{\left(\frac{1}{\lambda}\sqrt{\frac{\pi}{\beta}}\right)^n}{\sqrt{n}}=\frac{1}{\lambda}\sqrt{\frac{\beta}{\pi}}
  \textrm{Li}_{1/2}\left(\frac{1}{\lambda}\sqrt{\frac{\pi}{\beta}}\right)\ , \label{dlogF}
\end{equation}
where the polylogarithm function $\textrm{Li}_{s}(z)$ was introduced:
\begin{equation*}
  \textrm{Li}_{s}(z)=\sum_{n=1}^\infty \frac{z^n}{n^s},\quad |z|<1\ .
\end{equation*}
As mentioned at the beginning this derivation was assuming $\lambda>1$. When decreasing $\lambda$, one can see that the largest zero of $F$ leading to a singularity in (\ref{dlogF}) should obey:
\begin{equation*}
  \frac{1}{\lambda}\sqrt{\frac{\pi}{\beta}}=1\ , 
\end{equation*}
which yields for the leading eigenvalue
\begin{equation}
  \lambda_0(\beta)\simeq \sqrt{\frac{\pi}{\beta}}
\end{equation}
Using this approximation gives the mean energy per particle in the low temperature regime, i.e. the equipartition theorem, following (\ref{energytemperature}):
\begin{equation}
  h(\beta)\simeq \frac{1}{\beta},\quad \beta\to\infty\ .
\end{equation}
This coincides with the numerical estimate in Fig.~\ref{fig:Temp}.

\section{Mapping to sinh Gordon}\label{mapSinhGordon}

In this Section it is shown how to map the equation obtained in the continuum limit for $n=1$ with smoothed walls to the sinh Gordon field theory. Start with the Lagrangian in $(1+1)-$dimension:
\begin{equation}
  \mathcal{L}=\int_{-\infty}^{+\infty} \left[ \frac{1}{2} \left(\partial_t \phi\right)^2 - \frac{1}{2} \left(\partial_x \phi\right)^2 - V_2(\phi) \right] \ud x,
\end{equation}
where the potential $V_2(x)$ is given by (\ref{smoothwalls2}). The Euler-Lagrange equation is then:
\begin{equation}
  \partial_t^2 \phi - \partial_x^2 \phi = e^{\alpha \phi} - e^{-\alpha(x-1)} \label{EL_2walls}
\end{equation}
First change the unknown function:
$$ \phi(x,t) \mapsto \varphi(x,t)=\alpha\left( \phi(x,t) - \frac{1}{2} \right) \ ,$$
so that the field equation (\ref{EL_2walls}) becomes now
$$  \partial_t^2 \varphi - \partial_x^2 \varphi + 2\alpha e^{-\alpha/2} \sinh \varphi = 0 \ .$$
Then by rescaling the spacetime coordinates
\begin{equation}
  x\mapsto \xi=\sqrt{2\alpha e^{-\alpha/2}} x,\quad t \mapsto \tau =\sqrt{2\alpha e^{-\alpha/2}} t\ ,
\end{equation}
one gets the standard sinh Gordon field equation
\begin{equation}
  \label{SinhGordon}
  \partial_\tau^2 \varphi - \partial_\xi^2 \varphi + \sinh \varphi = 0
\end{equation}
\bibliographystyle{apsrev4-1}
\bibliography{bibi}
\end{document}